\documentclass[english,11pt,a4paper]{article}
\usepackage[T1]{fontenc}
\usepackage[latin9]{inputenc}
\usepackage{amsmath}
\usepackage{amsthm}
\usepackage{amssymb}
\usepackage{graphicx}
\usepackage{wasysym}

\makeatletter
\numberwithin{equation}{section}

\usepackage{jheppub}
\usepackage{slashed}
\usepackage[compat=1.1.0]{tikz-feynhand}
\usepackage{physics}
\usepackage{bm}
\usepackage{simpler-wick}
\usepackage{fontawesome5}
\allowdisplaybreaks

\makeatother

\usepackage{babel}
\begin{document}
\title{ A comprehensive calculation of the Primakoff process and
the solar axion flux}

\abstract{The Primakoff process plays a crucial role in axion production
in astrophysical environments and laboratories. Given the rising interest
in axion physics and many on-going experimental activities, we conduct
a comprehensive calculation of this process and carefully examine
several  aspects that have been neglected in the literature. In particular,
our calculation is valid for axions with significantly large masses,
which would be of importance to axion searches utilizing crystal and
liquid xenon detectors. We present the most updated calculation of
the Primakoff solar axion flux, with a simple parametrization that
is applicable to a broad range of axion masses up to a few tens of
keV. Our code is publicly available at GitHub \href{https://github.com/Fenyutanchan/Solar-Axion-Primakoff-Flux.git}{\faGithub}.}

\author[a, b]{Quan-feng Wu}
\author[a]{Xun-Jie Xu}
\affiliation[a]{Institute of High Energy Physics, Chinese Academy of Sciences, Beijing 100049, China}
\affiliation[b]{University of Chinese Academy of Sciences, Beijing 100049, China}
\emailAdd{wuquanfeng@ihep.ac.cn}
\emailAdd{xuxj@ihep.ac.cn}  
\preprint{\today}  
\maketitle

\section{Introduction}

The Primakoff process~\cite{Primakoff:1951iae}, originally proposed
for photoproduction of the $\pi^{0}$ meson in a nuclear electric
field,  currently plays a crucial role in the active  field of searching
for axions, a type of new light pseudoscalars well motivated by the
strong CP problem~\cite{Peccei:1977hh,Peccei:1977ur,Weinberg:1977ma,Wilczek:1977pj}.
In stellar environments,  axions could be abundantly produced via
the Primakoff process, leading to various  interesting astrophysical
consequences~\cite{Raffelt1996,Caputo:2024oqc}. In particular,
solar axions might be detectable in laboratories via the inverse Primakoff
process,  motivating a number of experimental searches~\cite{SOLAX:1997lpz,CAST:2007jps,CDMS:2009fba,Armengaud:2013rta,Majorana:2022bse}.

Given the importance of the Primakoff process in axion studies and
the actively on-going experimental searches, we believe  it is timely
to revisit this process and carefully examine several  aspects that
have been neglected in the literature. For instance, the simplest
Primakoff cross section of a photon scattering off a nucleus in the
vacuum, obtained  by treating the electric field around the  nucleus
as a classical field, cannot fully take into account the recoil, spin,
and magnetic moment of the nucleus. The widely-adopted formula for
the Primakoff production of axions in plasma with Debye-H\"{u}ckel
screening included, first derived in Ref.~\cite{Raffelt:1985nk},
is not applicable to massive axions if the mass is significant.\footnote{An earlier study \cite{Dicus:1978fp} considered the production of
heavy (keV to sub-MeV) axions in red giants, in which the mass effect
was important and taken into account, but the screening effect was
not included. } Although the axion mass effect is negligible in helioscope experiments
like CAST which quickly loses sensitivity  above a few eV~\cite{CAST:2017uph},
it could be of importance to experiments utilizing  crystal~\cite{SOLAX:1997lpz,CDMS:2009fba,Armengaud:2013rta,Majorana:2022bse}
or liquid xenon~\cite{Abe:2012ut,XENON100:2014csq,Gao:2020wer,Dent:2020jhf}
detectors if the search range extends to keV or higher. 

In this work, we present a comprehensive calculation of the Primakoff
process, aiming at providing a revised solar neutrino flux valid in
the full mass range.  Such a calculation would be particularly useful
for phenomenological studies on solar axions~\cite{Dafni:2018tvj,Jaeckel:2019xpa,Banerjee:2019xuy,Bhusal:2020bvx,DiLuzio:2020jjp,Dent:2020jhf,Gao:2020wer,Cai:2020bhd,OHare:2020wum,Guarini:2020hps,Bastero-Gil:2021oky,DeRocco:2022jyq,Lucente:2022esm,Beaufort:2023zuj,Hoof:2023jol}.
We start with a rigorous treatment of the kinematics and an exact
evaluation of the Feynman diagram of Primakoff scattering in the vacuum~\cite{Aloni:2019ruo}.
 Then step by step, we take into account more factors such as the
magnetic moment and spin of the target particle, the difference between
scalar and pseudoscalar, and Debye-H\"{u}ckel screening, and discuss
their impact thoroughly. We also derive the coherent photon-axion
conversion rate in a magnetic field in the quantum field theory (QFT)
approach, which not only helps to clarify subtitles arising from the
mismatch between the incoming and outgoing momenta, but also allows
us to obtain the conversion rate for non-relativistic axions.  This
could be important to axion dark matter conversion in e.g. neutron
star magnetospheres~\cite{Pshirkov:2007st,Huang:2018lxq,Hook:2018iia,Safdi:2018oeu,Battye:2019aco,Leroy:2019ghm,Foster:2020pgt,Prabhu:2021zve,Witte:2021arp,Battye:2021yue,Millar:2021gzs,Foster:2022fxn,Noordhuis:2022ljw,Battye:2023oac,Noordhuis:2023wid,Caputo:2023cpv}. 

This paper is organized as follows. In Sec.~\ref{sec:General}, we
setup the notation used in this work and elucidate the correlations
among multiple kinematic variables.  In Sec.~\ref{sec:cross-section},
we present the detailed calculation of the Primakoff process and discuss
the impact of many previously neglected factors. Then in Sec.~\ref{sec:Solar-axion-flux}
we apply the calculation to the latest solar model and obtain a practically
useful expression for  the solar axion flux. The codes for the flux
calculation is publicly accessible via GitHub\footnote{{\faGithub} \url{https://github.com/Fenyutanchan/Solar-Axion-Primakoff-Flux.git} }.
Finally we conclude in Sec.~\ref{sec:Conclusions} and relegate some
details to the appendix.

\section{General setup\protect\label{sec:General}}

\subsection{Lagrangian}

The electromagnetic interaction of the axion is given by the following
effective Lagrangian:
\begin{equation}
\mathcal{L}\supset-\frac{g_{a\gamma}}{4}aF_{\mu\nu}\tilde{F}^{\mu\nu}\thinspace,\label{eq:L}
\end{equation}
where $g_{a\gamma}$ is a dimensional coupling constant, $a$ denotes
the axion field, $F_{\mu\nu}\equiv\partial_{\mu}A_{\nu}-\partial_{\nu}A_{\mu}$
is the electromagnetic field tensor, and $\tilde{F}$ is the Hodge
dual of the field strength tensor $F$, i.e., 
\begin{equation}
\tilde{F}^{\mu\nu}=\frac{1}{2}\varepsilon^{\mu\nu\rho\sigma}F_{\rho\sigma}\thinspace,\label{eq:-1}
\end{equation}
with the convention $\varepsilon^{0123}=-\varepsilon_{0123}=1$. 

While Eq.~\eqref{eq:L} can naturally arise from the Peccei-Quinn
solution to the strong CP problem~\cite{Peccei:1977hh,Peccei:1977ur,Weinberg:1977ma,Wilczek:1977pj},
it is also one of the simplest interaction terms\footnote{At dimension-three level, one could write down the operator $\phi A^{\mu}A_{\mu}$
where $\phi$ is a generic scalar. But this operator does not respect
the gauge symmetry of QED. At dimension-four level, operators like
$\phi^{*}A^{\mu}\partial_{\mu}\phi$ could appear but $\phi$ has
to be a charged scalar which, with a mass below typical collider energy
scales and a quantized charge, would be efficiently produced on colliders
via the Drell-Yang process and thus be ruled out. Hence we conclude
that the simplest operators of light scalars interacting with the
electromagnetic field should be at least dimension-five. } that couple a new light (pseudo)scalar to the electromagnetic field.
  Given that the electromagnetic field is highly manipulable, Eq.~\eqref{eq:L}
also serves as a phenomenologically interesting avenue for new physics
exploration.

In addition to the pseudoscalar interaction, one may also consider
a scalar coupled to the electromagnetic field:
\begin{equation}
\mathcal{L}\supset-\frac{g_{\phi\gamma}}{4}\phi F_{\mu\nu}F^{\mu\nu}\thinspace,\label{eq:L-1}
\end{equation}
where $\phi$ is a CP-even scalar with a similar coupling $g_{\phi\gamma}$.
  Although in this work we mainly focus on the pseudoscalar case,
our results can be readily applied to the scalar case, as the difference
is negligible if particle spins or polarizations are averaged out---see
Secs.~\ref{subsec:P-vs-S} and \ref{subsec:coh-mag-fields}.  

From Eqs.~\eqref{eq:L} and ~\eqref{eq:L-1}, it is straightforward
to write down the Feynman rule of the $a$-$A$-$A$ (or $\phi$-$A$-$A$)
vertex:  

\begin{equation}
    \begin{tikzpicture}[baseline=(v.base)]
        \begin{feynhand}
            \vertex (v) at (0, 0);
            \vertex (a) at (-1, 0);
            \vertex (gamma-1) at (.5, .87) {$\mu$};
            \vertex (gamma-2) at (.5, -.87) {$\nu$};

            \propag [sca, mom=$p$] (a) to (v);
            \propag [pho, mom=$k$] (gamma-1) to (v);
            \propag [pho, mom=$k'$] (gamma-2) to (v);
        \end{feynhand}
    \end{tikzpicture} =
\begin{cases}
{\rm i}g_{a\gamma}\varepsilon^{\mu\nu\rho\sigma}k_{\sigma}k'_{\rho} & (\text{pseudoscalar})\\[2mm]
{\rm i}g_{\phi\gamma}\left[\left(k\cdot k'\right)g_{\mu\nu}-k_{\nu}k'_{\mu}\right] & (\text{scalar})
\end{cases}\,.
\label{eq:feyn}
\end{equation} They lead to almost identical decay widths:
\begin{equation}
\Gamma_{a\to\gamma\gamma}=\frac{g_{a\gamma}^{2}m_{a}^{3}}{64\pi}\thinspace,\ \ \Gamma_{\phi\to\gamma\gamma}=\frac{g_{\phi\gamma}^{2}m_{\phi}^{3}}{64\pi}\thinspace.\label{eq:-86}
\end{equation}

In terms of electric and magnetic fields, $\left(\mathbf{E},\ \mathbf{B}\right)\equiv\left(-\boldsymbol{\nabla}A^{0}-\partial_{t}\mathbf{A},\ \boldsymbol{\nabla}\times\mathbf{A}\right)$,
  the interactions in Eqs.~\eqref{eq:L} and ~\eqref{eq:L-1} can
be written as 
\begin{align}
-\frac{1}{4}aF_{\mu\nu}\tilde{F}^{\mu\nu} & =\left(\mathbf{E}\cdot\mathbf{B}\right)a\thinspace,\label{eq:-37}\\
-\frac{1}{4}\phi F_{\mu\nu}F^{\mu\nu} & =\frac{1}{2}\left(|\mathbf{E}|^{2}-|\mathbf{B}|^{2}\right)\phi\thinspace,\label{eq:-38}
\end{align}
which will be used in the calculation of photon-axion conversion in
macroscopic magnetic fields.

\subsection{Kinematics}

We consider the Primakoff process, $\gamma+X\to a+X$, where the target
particle $X$ can be  an electron, a proton, or a heavier nucleus.
The masses of $X$ and $a$ are denoted by $m_{X}$ and $m_{a}$,
respectively. For $m_{a}=0$, it is elastic scattering with relatively
simple kinematics. For $m_{a}\neq0$ (e.g.,~$m_{a}$ being comparable
to the incoming photon energy),  the kinematics becomes more complicated,
as we shall elaborate below. 

\begin{figure}
\centering

\raisebox{1.5cm}{\includegraphics[width=0.3\textwidth]{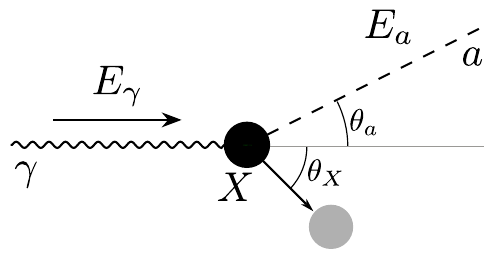}}\ \ \ \ \ \ \ \ \includegraphics[width=0.6\textwidth]{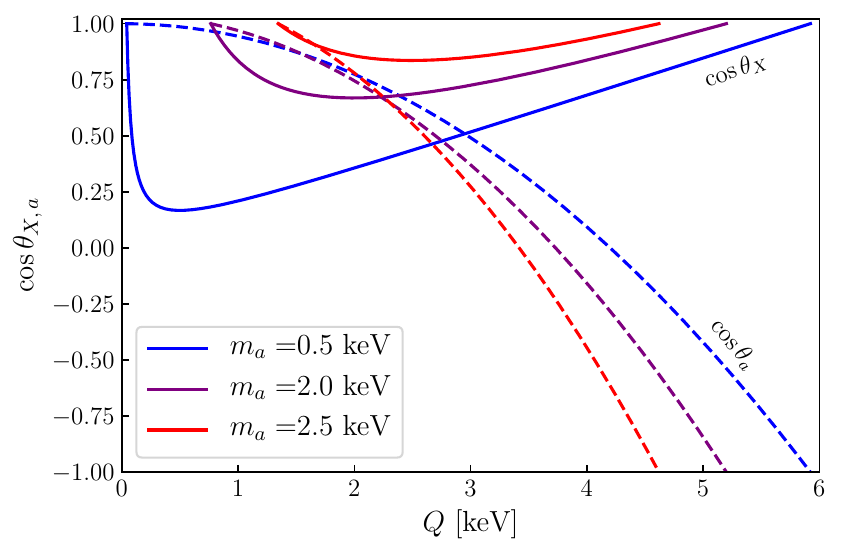}\caption{The kinematics of the Primakoff process. Left panel:  a schematic
representation of kinematic variables $\theta_{a}$ and $\theta_{X}$.
 Right panel: The variation of $\cos\theta_{a}$ (dashed lines) and
$\cos\theta_{X}$ (solid lines) as functions of the momentum transfer
$Q$, assuming $E_{\gamma}=3$ keV, $m_{a}\in\{0.5,\ 2.0,\ 2.5\}$
keV, and $m_{X}=0.511$ MeV. \protect\label{fig:kin}}
\end{figure}

We denote the incoming photon energy and the outgoing axion energy
by $E_{\gamma}$ and $E_{a}$, respectively. The angle between the
outgoing axion and the incoming photon is $\theta_{a}$. For the $X$
particle after scattering, we define a similar angle $\theta_{X}$
and denote its kinetic energy by $T_{X}$. These kinematic variables
are illustrated by the left panel of Fig.~\ref{fig:kin}. 

We further denote the four momenta of the incoming photon, the outgoing
axion, the $X$ particle before and after scattering by $p_{1}^{\mu}$,
$p_{3}^{\mu}$, $p_{2}^{\mu}$ and $p_{4}^{\mu}$, respectively. Without
loss of generality, we assume the following explicit forms for these
momenta:
\begin{align}
p_{1}^{\mu} & =(E_{\gamma},\ -E_{\gamma},\ 0,\ 0)\thinspace,\label{eq:-2}\\
p_{2}^{\mu} & =(m_{X},\ 0,\ 0,\ 0)\thinspace,\label{eq:-3}\\
p_{3}^{\mu} & =(E_{a},\ \mathbf{p}_{3})\thinspace,\label{eq:-4}\\
p_{4}^{\mu} & =(m_{X}+T_{X},\ \mathbf{p}_{4})\thinspace.\label{eq:-5}
\end{align}
The momentum transferred from the photon to the $X$ particle, $q^{\mu}\equiv p_{1}^{\mu}-p_{3}^{\mu}=p_{4}^{\mu}-p_{2}^{\mu}$,
is spacelike, $q^{2}=q\cdot q<0$. Hence we define 
\begin{equation}
Q\equiv\sqrt{-q^{2}}\thinspace,\label{eq:-11}
\end{equation}
which quantifies the momentum transfer. 

Next, we employ the Mandelstam variables: $s\equiv(p_{1}+p_{2})^{2}$,
$t\equiv(p_{1}-p_{3})^{2}$, and $u\equiv(p_{1}-p_{4})^{2}$.  The
$s$ variable can be computed directly from Eqs.~\eqref{eq:-2} and
\eqref{eq:-3}.  The $t$ variable can be computed from $t=(p_{2}-p_{4})^{2}=p_{2}^{2}-2p_{2}\cdot p_{4}+p_{4}^{2}$
with $p_{2,4}^{2}=m_{X}^{2}$ and $p_{2}\cdot p_{4}=m_{X}(m_{X}+T_{X})$.
The $u$ variable can be obtained by $\Sigma_{m^{2}}-s-t$ where $\Sigma_{m^{2}}$
denotes the sum of squares of all masses in the initial and final
states.  In terms of $T_{X}$, the Mandelstam variables are given
by
\begin{align}
s & =m_{X}(2E_{\gamma}+m_{X})\thinspace,\label{eq:-15}\\
t & =-2m_{X}T_{X}\thinspace,\label{eq:-16}\\
u & =m_{X}^{2}-2E_{\gamma}m_{X}+2m_{X}T_{X}+m_{a}^{2}\thinspace.\label{eq:-17}
\end{align}

The angles $\theta_{a}$ and $\theta_{X}$ can be obtained by computing
$p_{1}\cdot p_{3}$ and $p_{1}\cdot p_{4}$ using Eqs.~\eqref{eq:-4}
and \eqref{eq:-5}, and then comparing them with $p_{1}\cdot p_{3}=\frac{1}{2}\left(p_{1}^{2}+p_{3}^{2}-t\right)$
and $p_{1}\cdot p_{4}=\frac{1}{2}\left(p_{1}^{2}+p_{4}^{2}-u\right)$:
\begin{align}
\cos\theta_{X} & =\frac{2T_{X}(E_{\gamma}+m_{X})+m_{a}^{2}}{2E_{\gamma}\sqrt{2m_{X}T_{X}+T_{X}^{2}}}\thinspace,\label{eq:-6}\\
\cos\theta_{a} & =\frac{2E_{\gamma}^{2}-2E_{\gamma}T_{X}-2m_{X}T_{X}-m_{a}^{2}}{2E_{\gamma}\sqrt{E_{\gamma}^{2}-2E_{\gamma}T_{X}+T_{X}^{2}-m_{a}^{2}}}\thinspace.\label{eq:-7}
\end{align}
For later use, we also present an analytical expression of $d\cos\theta_{a}/dT_{X}$:
\begin{equation}
\frac{d\cos\theta_{a}}{dT_{X}}=\frac{-2E_{\gamma}^{2}m_{X}+E_{\gamma}\left(2m_{X}T_{X}+m_{a}^{2}\right)+m_{a}^{2}(2m_{X}+T_{X})}{2E_{\gamma}\left(E_{\gamma}^{2}-2E_{\gamma}T_{X}-m_{a}^{2}+T_{X}^{2}\right)^{3/2}}\thinspace.\label{eq:-20}
\end{equation}

The kinetic energy $T_{X}$ cannot vary arbitrarily, otherwise $\cos\theta_{X}$
and $\cos\theta_{a}$ in Eqs.~\eqref{eq:-6} and \eqref{eq:-7} could
exceed the allowed interval $[-1,\ 1]$. The kinematically allowed
range of $T_{X}$ is given by 
\begin{equation}
T_{\min}\leq T_{X}\leq T_{\max}\thinspace,\label{eq:-8}
\end{equation}
with 
\begin{align}
T_{\min} & =\frac{2E_{\gamma}^{2}m_{X}-m_{a}^{2}\left(E_{\gamma}+m_{X}\right)-E_{\gamma}\sqrt{4E_{\gamma}^{2}m_{X}^{2}-4m_{a}^{2}m_{X}\left(E_{\gamma}+m_{X}\right)+m_{a}^{4}}}{2m_{X}\left(2E_{\gamma}+m_{X}\right)}\thinspace,\label{eq:-9}\\
T_{\max} & =\frac{2E_{\gamma}^{2}m_{X}-m_{a}^{2}\left(E_{\gamma}+m_{X}\right)+E_{\gamma}\sqrt{4E_{\gamma}^{2}m_{X}^{2}-4m_{a}^{2}m_{X}\left(E_{\gamma}+m_{X}\right)+m_{a}^{4}}}{2m_{X}\left(2E_{\gamma}+m_{X}\right)}\thinspace.\label{eq:-10}
\end{align}
In the limit of $m_{a}\to0$, Eq.~\eqref{eq:-9} vanishes, as is expected
from elastic scattering. For small $m_{a}$, we obtain the following
expansion in $m_{a}$:
\begin{align}
T_{\min} & \approx\frac{m_{a}^{4}}{8E_{\gamma}^{2}m_{X}}+{\cal O}\left(m_{a}^{6}\right),\label{eq:-21}\\
T_{\max} & \approx\frac{2E_{\gamma}^{2}}{2E_{\gamma}+m_{X}}\left(1-\frac{E_{\gamma}+m_{X}}{2E_{\gamma}^{2}m_{X}}m_{a}^{2}\right)+{\cal O}\left(m_{a}^{4}\right).\label{eq:-22}
\end{align}

In the right panel of Fig.~\ref{fig:kin}, we vary $T_{X}$ from
its minimum to its maximum (the corresponding value of $Q$ is given
by $Q=\sqrt{2m_{X}T_{X}}$) and  plot the variation of $\cos\theta_{X}$
and $\cos\theta_{a}$ according to Eqs.~\eqref{eq:-6} and \eqref{eq:-7}.
As the figure shows, when $Q$ increases from the kinematically allowed
minimum to the maximum,   $\cos\theta_{a}$ decreases monotonically
from $1$ to $-1$, while $\cos\theta_{X}$ first decreases and then
increases, with a minimum appearing in the middle. 

The allowed interval of $T_{X}$ (or $Q$) shrinks as $m_{a}$ increases.
From Eqs.~\eqref{eq:-6} and \eqref{eq:-7} we obtain  that the interval
vanishes (i.e., $T_{\min}=T_{\max}$) when $m_{a}$ reaches
\begin{equation}
m_{a}^{{\rm max}}=\sqrt{2E_{\gamma}m_{X}+m_{X}^{2}}-m_{X}\approx E_{\gamma}-\frac{E_{\gamma}^{2}}{2m_{X}}+{\cal O}\left(\frac{E_{\gamma}^{3}}{m_{X}^{3}}\right).\label{eq:-12}
\end{equation}
Eq.~\eqref{eq:-12} implies that the maximal axion mass allowed by
the kinematics is slightly below the photon energy $E_{\gamma}$,
assuming $E_{\gamma}/m_{X}\ll1$.

\section{The cross section of the Primakoff process\protect\label{sec:cross-section}}

In this section, we present our calculations of the Primakoff process
at multiple levels starting from the simplest scenario where only
the electric charge form factor of the target particle is considered,
with more factors to be taken into account in subsequent subsections.
Whenever possible, we compare our results with known results in the
literature.  

\subsection{Scattering off electric charges\protect\label{subsec:electric}}

\begin{figure}[htpb]
\centering \begin{tikzpicture}
        \begin{feynhand}
            \vertex (photon) at (-2, 2) {$\gamma$};
            \vertex (a) at (2, 2) {$a$};
            \vertex (X-in) at (-2, 0) {$X$};
            \vertex (X-out) at (2, 0) {$X$};
            \vertex [NEblob, scale=.5] (XEM) at (0, 0) {};
            \vertex (aEM) at (0, 2);

            \propag [pho, mom=$p_1$] (photon) to (aEM);
            \propag [sca, mom=$p_3$] (aEM) to (a);

            \propag [fer, very thick, mom'=$p_2$] (X-in) to (XEM);
            \propag [fer, very thick, mom'=$p_4$] (XEM) to (X-out);

            \propag [photon, mom=$q$] (aEM) to [edge label'=$\gamma^*$] (XEM);
        \end{feynhand}
    \end{tikzpicture} \caption{Primakoff production of the axion.}
\label{fig:Primakoff-process-pseudo-scalar}
\end{figure}
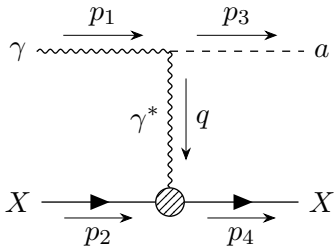

We first consider that $X$ is an electrically charged fermion interacting
with the photon via
\begin{equation}
{\cal L}\supset e\overline{\psi_{X}}F_{1}(q^{2})\gamma^{\mu}A_{\mu}\psi_{X}\thinspace,\label{eq:-13}
\end{equation}
where $\psi_{X}$ denotes the Dirac spinor of $X$ and $F_{1}(q^{2})$
is the charge form factor of $X$. The Feynman diagram of $\gamma+X\to a+X$
is presented in Fig.~\ref{fig:Primakoff-process-pseudo-scalar}.
Given the photon-axion vertex in Eq.~\eqref{eq:feyn}, the squared
matrix element of this process reads:
\begin{equation}
|{\cal M}|^{2}=\frac{1}{2}\cdot\frac{1}{2}\cdot\left|g_{a\gamma}\epsilon_{\mu}\varepsilon^{\mu\nu\alpha\beta}p_{1\alpha}q_{\beta}\frac{1}{q^{2}}eF_{1}\overline{u_{4}}\gamma_{\nu}u_{2}\right|^{2}\thinspace,\label{eq:-14}
\end{equation}
where $\epsilon$, $u_{4}$ and $u_{2}$ denote the initial photon,
final and initial fermion states, respectively. The two $1/2$ factors
account for spin and polarization average. 

Given the squared matrix element, the differential cross section is
computed by~\cite{Zyla:2020zbs}

\begin{equation}
\frac{d\sigma}{dt}=\frac{|{\cal M}|^{2}}{64\pi s|\mathbf{p}_{\text{1cm}}|^{2}}\thinspace,\label{eq:-18}
\end{equation}
where $s|\mathbf{p}_{\text{1cm}}|^{2}=(p_{1}\cdot p_{2})^{2}-p_{1}^{2}p_{2}^{2}=(s-m_{X}^{2})^{2}/4$. 

By applying the standard trace technology and contracting all Lorentz
indices, we obtain the explicit form of $|{\cal M}|^{2}$ expressed
in terms of the Mandelstam variables, among which $u$ can be replaced
by $\Sigma_{m^{2}}-s-t$ and $s$ by $m_{X}(2E_{\gamma}+m_{X})$ according
to Eq.~\eqref{eq:-15}. Then substituting the result of $|{\cal M}|^{2}$
into Eq.~\eqref{eq:-18}, we obtain 
\begin{equation}
\frac{d\sigma}{dt}=\frac{\alpha g_{a\gamma}^{2}F_{1}^{2}}{16}\left[-\frac{\left(t-m_{a}^{2}\right){}^{2}+4E_{\gamma}^{2}t}{2E_{\gamma}^{2}t^{2}}+\frac{m_{a}^{2}-t}{E_{\gamma}tm_{X}}-\frac{\left(t-m_{a}^{2}\right){}^{2}}{4E_{\gamma}^{2}tm_{X}^{2}}\right],\label{eq:-19}
\end{equation}
where the three terms in square brackets are proportional to $(1/m_{X})^{n}$
with $n=0$, $1$, and $2$. Consequently, the typical magnitudes
of these terms are in descending order from the left to the right.
However, in certain limits, the second term can be as large as the
first, resulting a cancellation between them, as we will see later.
We have checked that Eq.~\eqref{eq:-19} agrees with Eq.~(3) in Ref.~\cite{Aloni:2019ruo}
if the last term $\propto(1/m_{X})^{2}$ is omitted and the $F_{1}$
form factor is adjusted to the definition in Ref.~\cite{Aloni:2019ruo}.
In Sec.~\ref{subsec:spin0} we show that the $(1/m_{X})^{2}$ term
 should be absent for spin-0 target particles. 

\begin{figure}
\centering

\includegraphics[width=0.49\textwidth]{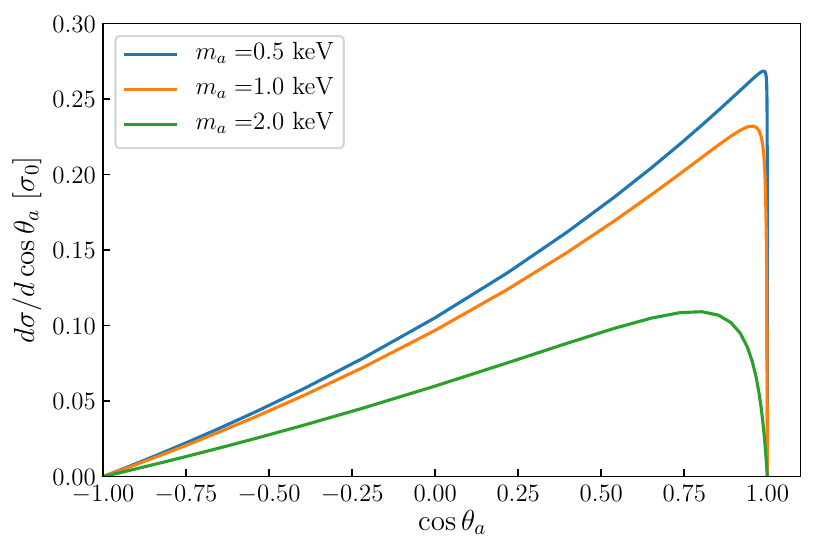}\includegraphics[width=0.49\textwidth]{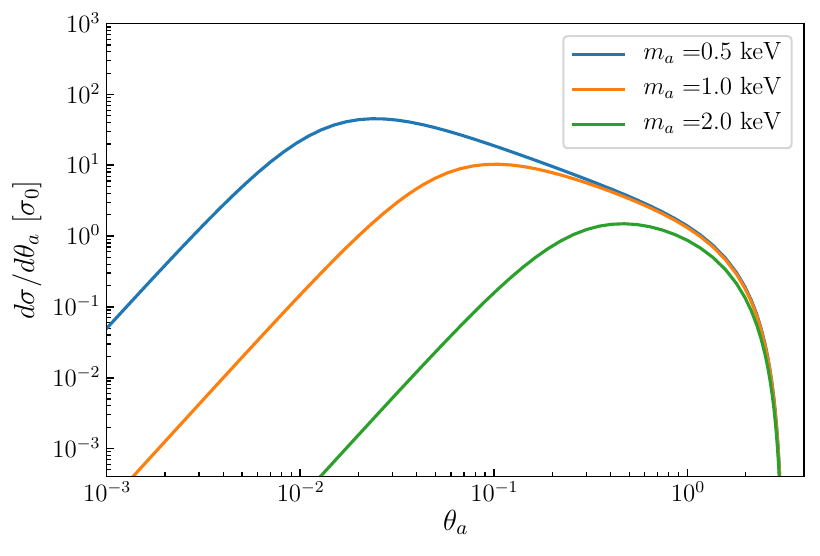}\caption{The differential cross sections $d\sigma/d\cos\theta_{a}$ (left panel)
and $d\sigma/d\theta_{a}$ (right panel) for $m_{a}\in\{0.5,\ 1,\ 2\}$
keV, $E_{\gamma}=3$ keV, and $m_{X}=m_{e}$. The unit $\sigma_{0}$
is defined in Eq.~\eqref{eq:-24}.\protect\label{fig:cross-section}}
\end{figure}

Eq.~\eqref{eq:-19} can be further written in terms of $T_{X}$: 

\begin{equation}
\frac{d\sigma}{dT_{X}}=\frac{\alpha g_{a\gamma}^{2}F_{1}^{2}}{8E_{\gamma}}\cdot\frac{T'_{X}}{T_{X}}\cdot J\thinspace,\ \ J\equiv\frac{2E_{\gamma}^{2}T_{X}-m_{X}T'_{X}{}^{2}}{2E_{\gamma}T_{X}T'_{X}}-1+\frac{T'_{X}}{2E_{\gamma}}\thinspace,\label{eq:-19-1}
\end{equation}
where $T'_{X}=T_{X}+m_{a}^{2}/(2m_{X})$. The differential cross
sections can also be written into the form of $d\sigma/d\cos\theta_{a}$
or $d\sigma/d\theta_{a}$, by multiplying proper Jacobian $dT_{X}/d\cos\theta_{a}$
or $dT_{X}/d\theta_{a}$---see Eq.~\eqref{eq:-20} for the analytical
expression. In Fig.~\ref{fig:cross-section}, we select a few representative
values of $m_{a}$ and present $d\sigma/d\cos\theta_{a}$ and  $d\sigma/d\theta_{a}$
as functions of  $\cos\theta_{a}$ and $\theta_{a}$ respectively. 

From Eq.~\eqref{eq:-19}, we obtain the total cross section:
\begin{equation}
\sigma_{{\rm tot}}\equiv\int\frac{d\sigma}{dT_{X}}dT_{X}\approx\sigma_{0}Q_{X}^{2}\left(L-1+K^{2}\right)+{\cal O}\left(m_{a}^{2}\right),\label{eq:-23}
\end{equation}
where we have taken $F_{1}^{2}(q^{2})\to Q_{X}^{2}$ with $Q_{X}$
the electric charge of $X$, and $(\sigma_{0},\ L,\ K)$ are defined
as
\begin{align}
\sigma_{0} & \equiv\frac{1}{8}\alpha g_{a\gamma}^{2}\thinspace,\label{eq:-24}\\
K & \equiv\frac{E_{\gamma}}{2E_{\gamma}+m_{X}}\thinspace,\label{eq:-25}\\
L & \equiv\log\left[\frac{16E_{\gamma}^{3}m_{X}K}{m_{a}^{4}}\right].\label{eq:-26}
\end{align}
The above result is obtained by expanding most quantities in $m_{a}$
except for a logarithmic part which is kept as the $L$ term. In the
limit of $m_{a}\to0$, the total cross section diverges, which is
a well-known feature of the Primakoff scattering in the vacuum~\cite{Raffelt:1985nk}.
In an electrically neutral medium, the divergence can be canceled
out after taking the Debye-H\"{u}ckel screening into account---see
Sec.~\ref{subsec:screening}. 

Within a rather broad range of $m_{a}$, $\sigma_{0}$ reflects the
typical magnitude of $\sigma_{{\rm tot}}$, since $L-1+K^{2}$ can
be roughly viewed as an ${\cal O}(1\sim10)$ quantity. In particular,
it implies that the total cross section is nearly independent of the
target mass, insofar as $E_{\gamma}\ll m_{X}$. Indeed, by taking
$K\approx E_{\gamma}/m_{X}\ll1$, we obtain $L\approx\log\left(16E_{\gamma}^{4}/m_{a}^{4}\right)$
and 
\begin{equation}
\sigma_{\text{tot}}(\gamma+e^{-}\to a+e^{-})-\sigma_{{\rm tot}}(\gamma+p\to a+p)\approx\sigma_{0}{\cal O}\left(\frac{E_{\gamma}}{m_{e}}\right).\label{eq:-27}
\end{equation}
Therefore, the  Primakoff cross sections with the electron and the
proton as target particles are almost identical, despite that their
masses are very different.\footnote{For comparison, the Thomson scattering cross section $8\pi\alpha^{2}/(3m_{X}^{2})$
is proportional to the inverse square of the target mass.} The fundamental reason for this is that the photon-to-axion conversion
rate mainly depends on the Coulomb potential around the target particle. 

There are two interesting limits of Eq.~\eqref{eq:-19}, namely the
forward and backward scattering limits, corresponding to $\theta_{a}\to0$
(or $T_{X}\to T_{\min}$) and $\theta_{a}\to\pi$ (or $T_{X}\to T_{\max}$).
In these limits, the $(1/m_{X})^{n}$ terms with $n=0$ and $1$ in
Eq.~\eqref{eq:-19} cancel out, for arbitrary non-vanishing $m_{a}$.
This can be seen by directly substituting $t=-2m_{X}T_{\min/\max}$,
with $T_{\min/\max}$ given by Eqs.~\eqref{eq:-9} and \eqref{eq:-10},
into Eq.~\eqref{eq:-19}.  Conversely, one can solve the equation
that equates the two terms, obtaining solutions that are exactly $T_{\min/\max}$
given by Eqs.~\eqref{eq:-9} and \eqref{eq:-10}. 

Since the first two terms cancel out in the forward and backward scattering
limits, we have 
\begin{equation}
\lim_{\theta_{a}\to0,\thinspace\pi}\ \frac{d\sigma}{dt}=\frac{\sigma_{0}Q_{X}^{2}}{2}\cdot\frac{\left(t-m_{a}^{2}\right){}^{2}}{4E_{\gamma}^{2}|t|m_{X}^{2}}\thinspace,\label{eq:-32}
\end{equation}
or
\begin{equation}
\lim_{\theta_{a}\to0,\thinspace\pi}\ \frac{d\sigma}{dT_{X}}=\frac{\sigma_{0}T'_{X}{}^{2}}{2E_{\gamma}^{2}T_{X}}\approx\frac{\sigma_{0}Q_{X}^{2}}{m_{X}}\times\begin{cases}
1-\frac{m_{a}^{2}}{2E_{\gamma}m_{X}}+{\cal O}\left(m_{a}^{4}\right) & (\theta_{a}\to0)\\
\frac{2E_{\gamma}m_{X}+m_{a}^{2}}{2E_{\gamma}m_{X}+4E_{\gamma}^{2}}+{\cal O}\left(m_{a}^{4}\right) & (\theta_{a}\to\pi)
\end{cases}\thinspace.\label{eq:-33}
\end{equation}
In the approximation of $\left(m_{a},\ E_{\gamma}\right)/m_{X}\ll1$,
the limit is simply given by $\sigma_{0}Q_{X}^{2}/m_{X}$.

\subsection{Including nuclear magnetic moments\protect\label{subsec:magnetic}}

The effective interaction of the proton with the photon can not be
fully described by Eq.~\eqref{eq:-13}. In fact, the proton has a
significantly large magnetic moment, which together with the large
magnetic moment of the neutron reveals that they are not fundamental
fermions and should  possess internal structures. 

To include the magnetic moment of the target particle into our calculation,
we extend Eq.~\eqref{eq:-13} to
\begin{equation}
{\cal L}\supset e\overline{\psi_{X}}\left[F_{1}(q^{2})\gamma^{\mu}+F_{2}(q^{2})\frac{i\sigma^{\mu\nu}q_{\nu}}{2m_{X}}\right]A_{\mu}\psi_{X}\thinspace,\label{eq:-13-1}
\end{equation}
which contains the most general form factors responsible for electromagnetic
interactions of a spin-1/2 particle, assuming the conservation of
$P$ and $CP$ symmetry. 

For the proton and the neutron, the form factors in the $q^{2}\to0$
limit are
\begin{align}
\text{proton}:\  & F_{1}=1,\ F_{2}=\mu_{p}-1,\ \ \text{with}\ \ \mu_{p}\approx2.7928\thinspace,\label{eq:-28}\\
\text{neutron}:\  & F_{1}=0,\ F_{2}=\mu_{n},\ \ \text{with}\ \ \mu_{n}\approx-1.9130\thinspace,\label{eq:-29}
\end{align}
where $\mu_{p}$ and $\mu_{n}$ are proton and neutron magnetic moments
in units of nucleon magneton {[}$e/(2m_{p})\approx0.1052\ e\cdot\text{fm}${]}.
Heavy nuclei may also possess significantly large magnetic moments,
which can be readily included in our calculation. For the electron,
 one-loop radiative corrections give rise to $F_{2}\approx\alpha/(2\pi)$.

\begin{figure}
\centering

\includegraphics[width=0.49\textwidth]{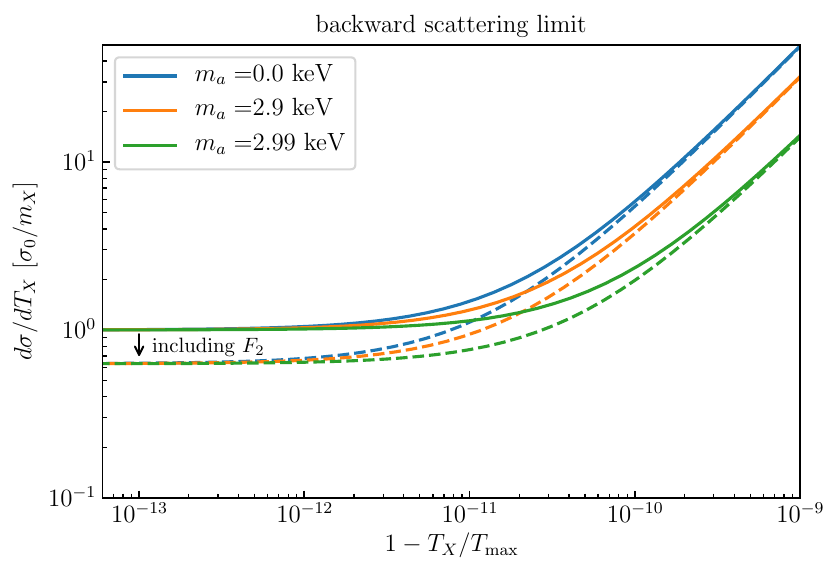}\includegraphics[width=0.49\textwidth]{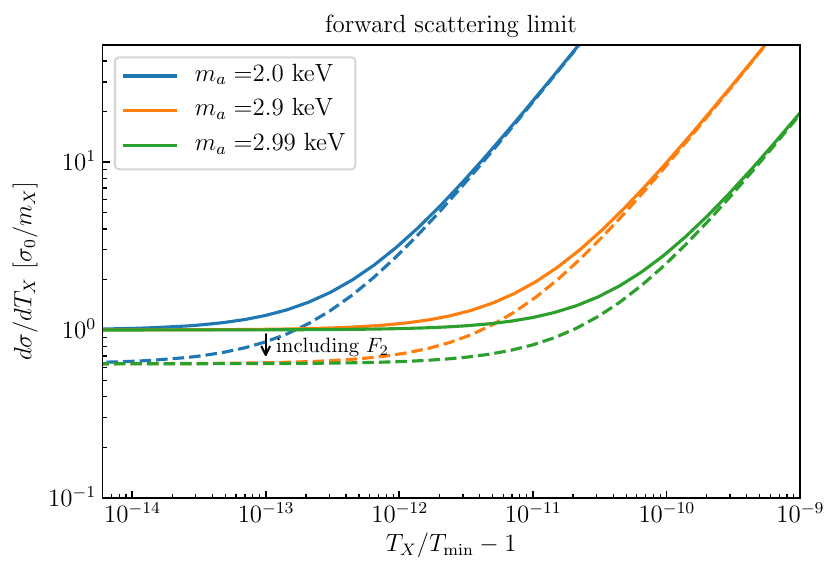}\caption{The effect of the proton magnetic moment on the Primakoff cross section.
The solid lines represent the cross section  including only the $F_{1}$
form factor---see Eq.~\eqref{eq:-19-1}. The dashed lines represent
the cross section including both $F_{1}$ and $F_{2}$ form factors---see
Eqs.~\eqref{eq:-30} and \eqref{eq:-31}. Note that this effect is
significant only in the limit of $T_{X}\to T_{\max}$ (left panel)
or $T_{X}\to T_{\min}$ (right panel). In this figure, we assume $E_{\gamma}=3$
keV, $F_{2}=1.793$, and $m_{X}=931$ MeV. \protect\label{fig:magnetic} }
\end{figure}

Using the effective electromagnetic vertex in Eq.~\eqref{eq:-13-1},
we find that the contribution of $F_{2}$ to the differential cross
section can be accounted for by
\begin{equation}
\frac{d\sigma}{dT_{X}}\to\left(1+\delta_{T}\right)\frac{d\sigma}{dT_{X}}\thinspace,\label{eq:-30}
\end{equation}
where
\begin{equation}
\delta_{T}\equiv\frac{T_{X}}{2JT'_{X}}\left[\frac{F_{2}^{2}}{F_{1}^{2}}\left(\frac{T'_{X}{}^{2}}{2E_{\gamma}T_{X}}+\frac{E_{\gamma}-T'_{X}}{m_{X}}\right)-\frac{F_{2}}{F_{1}}\frac{2T'_{X}{}^{2}}{E_{\gamma}T_{X}}\right],\label{eq:-31}
\end{equation}
with the $J$ factor defined in Eq.~\eqref{eq:-19-1}. Here the last
term $\propto F_{2}/F_{1}$ comes from the interference between $F_{1}$
and $F_{2}$. For electrically neutral (i.e. $F_{1}\to0$) particles
such as the neutron, Eqs.~\eqref{eq:-30} and \eqref{eq:-31} can
still be applied, with the notice that $\delta_{T}F_{1}^{2}$ is finite
in the $F_{1}\to0$ limit and the interference vanishes. 

The $\delta_{T}$ factor is only significant in the forward or backward
scattering limit. By taking $T_{X}\to T_{\min}$ or $T_{\max}$ (corresponding
to $\theta_{a}\to0$ or $\pi$) given by Eqs.~\eqref{eq:-9} and \eqref{eq:-10},
we obtain the following limit
\begin{equation}
\lim_{\theta_{a}\to0,\thinspace\pi}\ \delta_{T}=\frac{F_{2}}{F_{1}}\left(\frac{F_{2}}{F_{1}}-2\right),\label{eq:-34}
\end{equation}
which for the proton approximately equals to $-0.371$.

In Fig.~\ref{fig:magnetic}, we plot the cross sections with (dashed
lines) and without (solid lines) including the $F_{2}$ factor. 
As is expected, the $F_{2}$ factor reduces the Primakoff cross section
in the forward or backward scattering limit by about $37\%$.

\subsection{Target particles of spin 0, spin 1/2, ...\protect\label{subsec:spin0}}

In some circumstances, the target particle $X$ is not a spin-1/2
fermion and may possess lower or higher spins. For example, the $^{4}\text{He}$
nucleus is a spin-0 boson while the $^{2}\text{H}$ nucleus is a spin-$1$
boson. Heavier nuclei may possess higher spins. In general, the effect
of spin is small in low-energy scattering, roughly of the order of
$T'_{X}/E_{\gamma}\sim E_{\gamma}/m_{X}$.  Therefore, we only inspect
the difference between spin-0 and spin-1/2 target particles quantitatively. 

For a spin-0 particle, the electromagnetic vertex is formulated as
\begin{equation}
{\cal L}\supset ieF_{1}(q^{2})A_{\mu}X^{*}\partial^{\mu}X+{\rm h.c.}\thinspace,\label{eq:-13-2}
\end{equation}
where $X$ is a complex scalar. Eq.~\eqref{eq:-13-2} can be derived
from the theory of scalar QED, which would also give rise to another
interaction $A_{\mu}A_{\nu}\partial^{\mu}X^{*}\partial^{\nu}X$, as
requested by the gauge invariance. This interaction is not relevant
to our analysis but could be of importance to other axion production
processes. 

With the interaction in Eq.~\eqref{eq:-13-2}, Eq.~\eqref{eq:-14}
should be modified by replacing $eF_{1}\overline{u_{4}}\gamma_{\nu}u_{2}$
$\to eF_{1}(p_{2}+p_{4})_{\nu}$ and $\frac{1}{2}\cdot\frac{1}{2}\to\frac{1}{2}$
due to the absence of spin average for the  initial $X$ particle.
By repeating the calculation for the spin-0 case, we find that the
difference is
\begin{equation}
\left.\frac{d\sigma}{dT_{X}}\right|_{{\rm spin\thinspace1/2}}-\left.\frac{d\sigma}{dT_{X}}\right|_{{\rm spin\thinspace0}}=\frac{\alpha g_{a\gamma}^{2}F_{1}^{2}T'_{X}{}^{2}}{16E_{\gamma}^{2}T_{X}}\thinspace,\label{eq:-35}
\end{equation}
which implies that the last term in the $J$ factor of Eq.~\eqref{eq:-19-1}
exactly corresponds to the difference between spin-0 and 1/2. 

Note that this difference is suppressed by the heavy mass of the target,
since typically we have $T_{X}'\sim T_{X}\sim E_{\gamma}^{2}/m_{X}$
and hence $T'_{X}{}^{2}/E_{\gamma}T_{X}\sim E_{\gamma}/m_{X}$. More
generally, all spin effects should be negligible in the heavy mass
limit. In particular, both the spin-0 and spin-1/2 cross sections
in this limit can be written as 
\begin{equation}
\lim_{m_{X}\to\infty}\ \frac{d\sigma}{d\cos\theta_{a}}=\frac{1}{4}\alpha g_{a\gamma}^{2}F_{1}^{2}\frac{|\mathbf{p}_{1}\times\mathbf{p}_{3}|^{2}}{|\mathbf{q}|^{4}}\cdot\sqrt{1-\frac{m_{a}^{2}}{E_{\gamma}^{2}}}\thinspace,\label{eq:-36}
\end{equation}
which reproduces Eq.~(17) in Ref.~\cite{Raffelt:1985nk}, except
for an additional factor $\sqrt{1-m_{a}^{2}/E_{\gamma}^{2}}$ which
can be omitted when $m_{a}\ll E_{\gamma}$.

\subsection{Pseudoscalar versus scalar \protect\label{subsec:P-vs-S}}

So far our calculation has been focused on the axion which is a pseudoscalar.
For a scalar coupled to the photon via Eq.~\eqref{eq:L-1} {[}see
also Eq.~\eqref{eq:-38}{]}, the resulting scattering amplitude is
apparently quite different. Interestingly, after squaring the amplitude
and averaging spins and polarization, the final result is exactly
identical to the pseudoscalar case (here and in subsequent discussions
on the pseudoscalar-scalar difference, we assume $g_{a\gamma}=g_{\phi\gamma}$).
This can be verified by straightforwardly repeating the previous calculations
for the scalar case with the Feynman rule in Eq.~\eqref{eq:feyn}. 

More generally, one can prove that this is also true for the arbitrary
diagrams containing a (pseudo)scalar-photon-photon vertex with one
of the photons on-shell, insofar as only the spin-averaged squared
amplitude is concerned. For example, the decay widths of $\phi\to2\gamma$
and $a\to2\gamma$ are identical. 

For the Primakoff process, we can decompose the amplitude as
\begin{equation}
{\cal M}=\epsilon_{\mu}{\cal T}^{\mu\nu}{\cal X}_{\nu}\thinspace,\label{eq:-39}
\end{equation}
where ${\cal X}_{\nu}$ denotes the part of the amplitude that depends
on the properties of the $X$ particle, while $\epsilon_{\mu}{\cal T}^{\mu\nu}$
represents the remaining part, including the photon polarization vector
$\epsilon_{\mu}$. For the pseudoscalar and scalar cases, the ${\cal T}^{\mu\nu}$
part reads:
\begin{align}
{\cal T}_{\text{pseudo}}^{\mu\nu} & =g_{a\gamma}\varepsilon^{\mu\nu\alpha\beta}p_{1\alpha}q_{\beta}\frac{1}{q^{2}}\thinspace,\label{eq:-40}\\
{\cal T}_{\text{scalar}}^{\mu\nu} & =g_{\phi\gamma}\left(p_{1}^{\nu}q^{\mu}-g^{\mu\nu}p_{1}\cdot q\right)\frac{1}{q^{2}}\thinspace.\label{eq:-41}
\end{align}
When squaring the amplitude, we encounter the evaluation of ${\cal X}_{\nu}{\cal X}_{\rho}^{*}$,
which after averaging the spin (if exists) can always be written as
\begin{equation}
{\cal X}_{\nu}{\cal X}_{\rho}^{*}\xrightarrow{\text{spin average}}Ap_{2\rho}p_{4\nu}+Bp_{2\nu}p_{4\rho}+Cm_{X}^{2}g_{\rho\nu}+D\varepsilon^{\nu\rho\alpha\beta}p_{2\alpha}p_{4\beta},\label{eq:-42}
\end{equation}
where $A$, $B$, $C$, and $D$ represent quantities without Lorentz
structures. Eq.~\eqref{eq:-42} is the most general Lorentz-invariant
form one can write down for ${\cal X}_{\nu}{\cal X}_{\rho}^{*}$.

Combing Eqs.~\eqref{eq:-42}, \eqref{eq:-40},  \eqref{eq:-41}, and
\eqref{eq:-39}, we obtain
\begin{equation}
|{\cal M}_{\text{pseudo}}|^{2}-|{\cal M}_{\text{scalar}}|^{2}=m_{\gamma}^{2}g_{a\gamma}^{2}\frac{4m_{X}^{2}(A+B+3C)-t(A+B)}{8t}\thinspace,\label{eq:-43}
\end{equation}
where we have assumed a mass, $m_{\gamma}\equiv\sqrt{p_{1}^{\mu}p_{1\mu}}$,
for the incoming photon. Eq.~\eqref{eq:-43} implies that the difference
could be nonzero only when the incoming photon has a different dispersion
relation from the one in the vacuum. In  plasma, the plasmon mass
could play the role of $m_{\gamma}$ here but since it is typically
well below the photon energy, the resulting difference is small.

\subsection{Debye-H\"{u}ckel screening \protect\label{subsec:screening}}

In a medium, the Primakoff process may have a significantly altered
cross section.  The most prominent medium effect is the so-called
Debye-H\"{u}ckel screening~\cite{Raffelt:1985nk,Raffelt1996}. Since
each electric charge in a neutral medium is in average surrounded
by a charge cloud of the opposite sign, which at a sufficiently large
scale is capable to cancel the Coulomb potential generated by the
central charge, the effective Coulomb potential contributing to Primakoff
scattering should have a finite extent, beyond which its contribution
becomes negligible. Consequently, Primakoff scattering at low-momentum
transfer should take a reduced charge form factor. 

According to Ref.~\cite{Raffelt:1985nk}, Debye-H\"{u}ckel screening
for the Primakoff process in a thermal plasma reduces the $F_{1}$
form factor by
\begin{equation}
F_{1}^{2}(q^{2})\to F_{1}^{2}(q^{2})\frac{|q^{2}|}{|q^{2}|+\kappa_{s}^{2}}\thinspace,\label{eq:-66}
\end{equation}
with
\begin{equation}
\kappa_{s}^{2}\equiv\frac{4\pi\alpha}{T}\left(\sum_{X}Q_{X}^{2}n_{X}\right).\label{eq:-67}
\end{equation}
Here $T$ is the temperature of the plasma, and  $n_{X}$ denotes
 the number density of  $X$. For the summation in Eq.~\eqref{eq:-67},
one should take into account all electrically charged particles in
the medium, including electrons, protons and heavier nuclei, such
that $\sum_{X}Q_{X}n_{X}=0$. For the solar medium, $\kappa_{s}$
is about $9$ keV in the solar center and decreases in outer regions,
while $(\kappa_{s}/T)^{2}\approx12$ is approximately constant~\cite{Raffelt:2006cw}.

\begin{figure}
\centering

\includegraphics[width=0.7\textwidth]{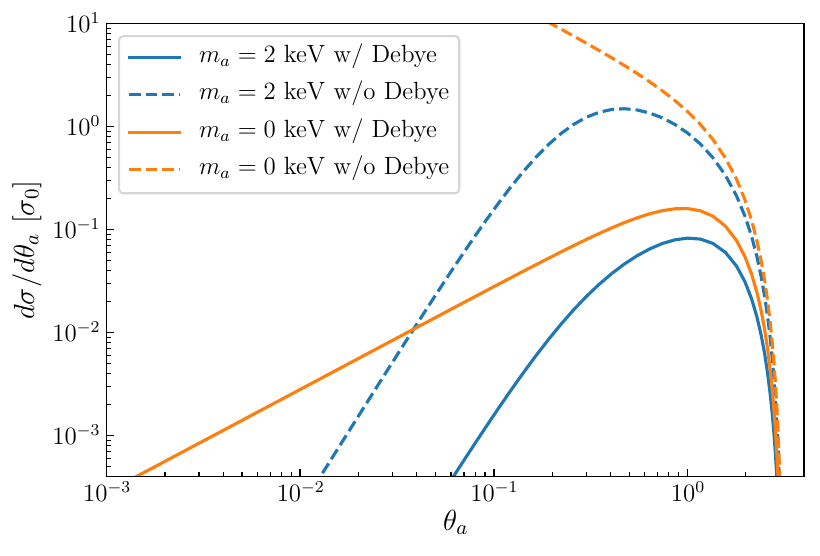}\caption{Primakoff cross sections with (solid) and without (dashed) Debye-H\"{u}ckel
screening. \protect\label{fig:Debye} }
\end{figure}

Fig.~\ref{fig:Debye} shows the impact of Debye-H\"{u}ckel screening
on the Primakoff cross section, assuming $E_{\gamma}=3$ keV, $\kappa_{s}=8$
keV, $m_{a}\in\{0,\ 2\}$ keV, and the electron as the target particle.
For $m_{a}=0$, $d\sigma/d\theta_{a}$ in the vacuum diverges in the
$\theta_{a}\to0$ limit, while in a medium the divergence is canceled
by Debye-H\"{u}ckel screening. 

Including the screening effect, the total cross section is finite
--- for the full expression, see Appendix~\ref{sec:Full-expression}.
In particular, in the large $m_{X}$ limit, we obtain the following
analytical expression for the total cross section:
\begin{equation}
\lim_{m_{X}\to\infty}\sigma_{\text{tot}}=\frac{\alpha g_{a\gamma}^{2}Q_{X}^{2}}{8}\left[\frac{4E_{\gamma}^{2}+\kappa_{s}^{2}-2m_{a}^{2}}{4E_{\gamma}^{2}}\log\frac{\kappa_{s}^{2}+q_{+}^{2}}{\kappa_{s}^{2}+q_{-}^{2}}+\frac{m_{a}^{4}}{4E_{\gamma}^{2}\kappa_{s}^{2}}\log\frac{m_{a}^{4}+\kappa_{s}^{2}q_{-}^{2}}{m_{a}^{4}+\kappa_{s}^{2}q_{+}^{2}}-\xi\right],\label{eq:-76}
\end{equation}
where 
\begin{equation}
\xi\equiv\sqrt{1-\frac{m_{a}^{2}}{E_{\gamma}^{2}}}\thinspace,\ \ q_{\pm}^{2}\equiv2E_{\gamma}^{2}\left(1\pm\xi\right)-m_{a}^{2}\thinspace.\label{eq:-77}
\end{equation}
In the $m_{a}\to0$, Eq.~\eqref{eq:-76} reduces to
\begin{equation}
\lim_{m_{a}\to0}\lim_{m_{X}\to\infty}\sigma_{\text{tot}}=\frac{\alpha g_{a\gamma}^{2}Q_{X}^{2}}{8}\left[\left(1+\frac{\kappa_{s}^{2}}{4E_{\gamma}^{2}}\right)\log\left(1+\frac{4E_{\gamma}^{2}}{\kappa_{s}^{2}}\right)-1\right],\label{eq:-78}
\end{equation}
which agrees with Eq.~(29) in Ref.~\cite{Raffelt:1985nk}. 

\subsection{Coherent scattering with large-scale magnetic fields\protect\label{subsec:coh-mag-fields}}

 In Secs.~\ref{subsec:electric} and \ref{subsec:magnetic}, we
have computed the Primakoff scattering with the electric and magnetic
fields of a single target particle. Electric and  magnetic fields
at macroscopic scales can be formed from a large collection of particles
participating electromagnetic interactions, both in nature and in
laboratories. In the presence of a large-scale electromagnetic field,
the incoming photon can scatter coherently with the field such that
the photon-axion conversion rate may exhibit the oscillation phenomenon.

Since the persistence of large-scale electric fields is rare in nature
and the maintenance of strong electric fields in laboratories is more
energy-consuming,  here we only consider large-scale magnetic fields.
 Theoretically, due to electric--magnetic duality, the calculation
for magnetic fields can be easily adapted for electric fields. 

In the presence of a large-scale magnetic field $\mathbf{B}$, the
amplitude of the photon $\gamma$ scattering with $\mathbf{B}$ and
producing the axion $a$ reads
\begin{equation}
{\cal A}_{\gamma\to a}\propto g_{a\gamma}\langle\wick{\c1a|\c1a\c2{\mathbf{E}}\cdot\mathbf{B}|\c2\gamma}\rangle\,,\label{eq:wick-axion}
\end{equation}
where $a$ and $\mathbf{E}$ are quantized fields which are to be
contracted with the initial and final single-particle states. In a
consistent QFT treatment, $\mathbf{B}$ should also be  quantized,
but  due to the enormously large occupation number of quanta in the
background, $\mathbf{B}$ can be replaced by its classical field value. 

Eq.~\eqref{eq:wick-axion} is interpreted as coherent scattering of
the photon with the magnetic field. The scattering may happen at any
point in the region with nonzero field strength, and all possible
scattering amplitudes should be summed coherently, corresponding to
an integration over the space. The outgoing axion may not have exactly
the same momentum as the incoming photon, subjected to certain limitations
from coherency. In particular,  Eq.~\eqref{eq:wick-axion} also allows
for coherent backward scattering, which means that the outgoing axion
moves along the opposite direction as the incoming photon while still
maintaining coherence. We leave discussions on these issues to Appendix~\ref{sec:QFT-coh}.

Assuming that the magnetic background is static and distributes vertically,
one can prove (see also Appendix~\ref{sec:QFT-coh}) that the energy
conservation and the conservation of momentum along the direction
parallel to the magnetic field are respected, i.e., 
\begin{equation}
E_{\gamma}=E_{a}\thinspace,\ \ \mathbf{p}_{\gamma,\parallel}=\mathbf{p}_{a,\parallel}\thinspace,\ \ \mathbf{p}_{\gamma,\perp}\neq\mathbf{p}_{a,\perp}\thinspace,\label{eq:-68}
\end{equation}
where the subscripts ``$\perp$'' and ``$\parallel$'' denote
projections perpendicular and parallel to the magnetic field. For
simplicity, below we assume  $\mathbf{p}_{\gamma,\parallel}=\mathbf{p}_{a,\parallel}=0$,
i.e., $\mathbf{p}_{\gamma}$ is perpendicular to $\mathbf{B}$. 

In momentum space, the contraction in Eq.~\eqref{eq:wick-axion} leads
to
\begin{equation}
\wick{\c1{\mathbf{E}}|\c1\gamma}\rangle\to\mathbf{p}_{\gamma}\wick{\c1{A}^{0}|\c1\gamma\rangle}+E_{\gamma}\wick{\c1{\mathbf{A}}|\c1\gamma\rangle}=\mathbf{p}_{\gamma}\epsilon^{0}+E_{\gamma}\boldsymbol{\epsilon}\thinspace,\ \ \langle\wick{\c1{a}|\c1a}\to1\thinspace,\label{eq:-69}
\end{equation}
where $(\epsilon^{0},\ \boldsymbol{\epsilon})$ are the temporal and
spatial components of the polarization vector $\epsilon^{\mu}$. Therefore,
the amplitude is
\begin{equation}
{\cal A}_{\gamma\to a}\propto g_{a\gamma}E_{\gamma}\boldsymbol{\epsilon}\cdot\mathbf{B}=g_{a\gamma}E_{\gamma}B\cos\theta_{\epsilon}\,,\label{eq:-70}
\end{equation}
where $\theta_{\epsilon}$ is defined as the angle between $\boldsymbol{\epsilon}$
and $\mathbf{B}$. The $\mathbf{p}_{\gamma}\epsilon^{0}$ term in
Eq.~\eqref{eq:-69} does not contribute because $\epsilon^{0}=0$
for massless photons. Eq.~\eqref{eq:-70} implies that the photon-axion
conversion rate becomes maximal when $\boldsymbol{\epsilon}$ is parallel
to $\mathbf{B}$. 

For the scalar interaction in Eq.~\eqref{eq:-38}, we have obtained
a similar result:
\begin{equation}
{\cal A}_{\gamma\to\phi}\propto g_{\phi\gamma}\left(\mathbf{p}_{\gamma}\times\boldsymbol{\epsilon}\right)\cdot\mathbf{B}=g_{\phi\gamma}\left(\boldsymbol{\epsilon}\times\mathbf{B}\right)\cdot\mathbf{p}_{\gamma}=g_{\phi\gamma}E_{\gamma}B\sin\theta_{\epsilon}\,,\label{eq:-71}
\end{equation}
which implies that the conversion rate is maximal  when $\boldsymbol{\epsilon}$
is perpendicular to $\mathbf{B}$. 

Performing Fourier transform from momentum space to coordinate space
(see Appendix~\ref{sec:QFT-coh} for details), we obtain the full
expression of ${\cal A}_{\gamma\to a}$: 
\begin{equation}
{\cal A}_{\gamma\to a}=\frac{g_{a\gamma}E_{\gamma}B\cos\theta_{\epsilon}}{2\sqrt{p_{\gamma}p_{a}}}\frac{e^{i|\mathbf{q}|L}-1}{|\mathbf{q}|}\thinspace,\label{eq:-72}
\end{equation}
where $L$ denotes the length of the photon propagation in the $\mathbf{B}$
field, $p_{\gamma}=E_{\gamma}$, $p_{a}=\sqrt{E_{\gamma}^{2}-m_{a}^{2}}$,
and $|\mathbf{q}|=E_{\gamma}-\sqrt{E_{\gamma}^{2}-m_{a}^{2}}$. Squaring
the amplitude, we obtain the transition probability:
\begin{equation}
P_{\gamma\to a}=\frac{g_{a\gamma}^{2}B^{2}\cos^{2}\theta_{\epsilon}}{|\mathbf{q}|^{2}}\cdot\frac{E_{\gamma}}{p_{a}}\cdot\sin^{2}\left(\frac{|\mathbf{q}|L}{2}\right).\label{eq:-73}
\end{equation}
For the inverse process $a\to\gamma$, the transition probability
is the same---see discussions below Eq.~\eqref{eq:-64}. 

Eq.~\eqref{eq:-73} is to be compared with e.g. Eq.~(25) in Ref.~\cite{Caputo:2024oqc}
(see also Eq.~(16) in Ref.~\cite{CAST:2007jps} and Eq.~(16) in
Ref.~\cite{vanBibber:1988ge}). We note here that Eq.~\eqref{eq:-73}
remains valid even for non-relativistic axions, while previous results
are only applicable to relativistic axions. The difference is accounted
for by the $E_{\gamma}/p_{a}$ factor. 

\subsection{Summary}

\begin{figure}
\centering

\includegraphics[width=0.69\textwidth]{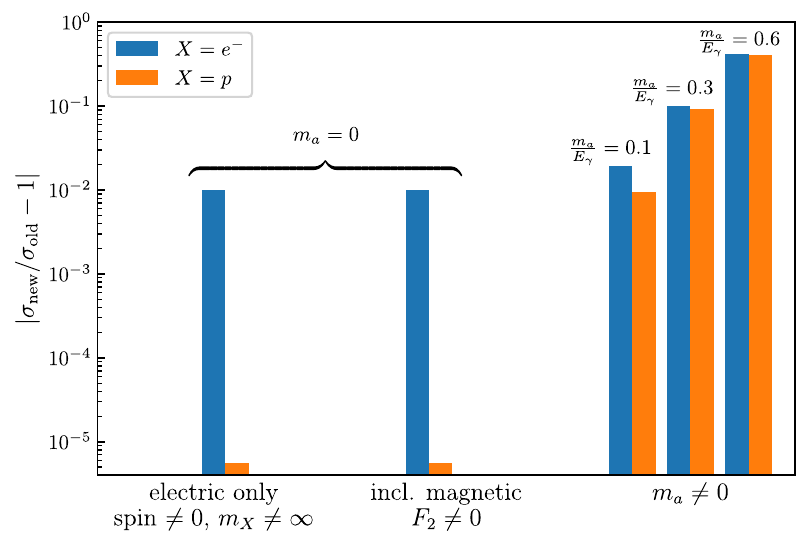}\caption{Deviations of the total cross section, denoted by $\sigma_{{\rm new}}$,
from the old one in Eq.~\eqref{eq:-78} which is denoted by $\sigma_{{\rm old}}$.
The bar heights indicate the magnitudes of corrections due to various
effects discussed in this work. In this figure, we take $E_{\gamma}=3$
keV and $\kappa_{s}=8$ keV. \protect\label{fig:correction}}
\end{figure}

In this section, we have comprehensively discussed various effects
that can cause corrections to the Primakoff cross section. In previous
studies, the most widely-adopted formula for the Primakoff cross section
is Eq.~\eqref{eq:-78}, which is obtained assuming that (i) the axion
is massless, and (ii) the target particle $X$ is infinitely heavy
($m_{X}\to\infty$) such that its spin and magnetic moment can be
neglected. Here we would like to discuss and compare the aforementioned
corrections to this widely-adopted cross section.  

For massless axions, the spin and magnetic moment  corrections crucially
depend on the finite mass $m_{X}$. This is shown in Fig.~\ref{fig:correction}
by the four bars on the left. The first two bars labeled ``electric
only'' with ``spin $\neq$ 0'' and ``$m_{X}\neq\infty$'' are
obtained by integrating Eq.~\eqref{eq:-19-1} over $T_{X}$, with
Eq.~\eqref{eq:-66} included, and comparing it with Eq.~\eqref{eq:-78}.
The next two bars labeled ``incl. magnetic'' and ``$F_{2}\neq0$''
show the corrections after further taking the $F_{2}$ factor into
account. Since  these effects would vanish in the limit of $m_{X}\to\infty$,
the correction for $X=e^{-}$ is much more significant than that for
$X=p$.

For massive axions, our new cross section can be significantly different
from the old one, as is shown by the bars labeled ``$m_{a}\neq0$''
in Fig.~\ref{fig:correction}. Here the correction due to the axion
mass is much larger than the corrections caused by the spin and magnetic
moment of the target particle. This correction has weak dependence
on the target particle mass, making the correction for $X=p$ slightly
lower than that for $X=e^{-}$. 

From Fig.~\ref{fig:correction}, we see that  for massive axions
with $m_{a}/E_{\gamma}\gtrsim0.1$, the correction due to finite $m_{a}$
is the dominant one. For $m_{a}/E_{\gamma}\ll0.1$ and $X=e^{-}$,
the spin and magnetic moment corrections become dominant, typically
of the order of $1\%$. This $\sim1\%$ correction cannot be reduced
by $m_{a}\to0$. Therefore, we conclude that the old formula in Eq.~\eqref{eq:-78}
is valid at $\sim1\%$ precision in the massless limit. If higher
precision is required or if the finite-$m_{a}$ correction is significant,
the cross section should include the relevant corrections.

\section{Solar axion flux \protect\label{sec:Solar-axion-flux}}

Given the crucial role the Sun has played in previous studies on axions
and the actively ongoing experimental searches,  we would like to
apply our results in Sec.~\ref{sec:cross-section} to the calculation
of solar axion flux\footnote{For axions emitted from other stellar environments, we refer to Ref.~\cite{Nguyen:2023czp}.
In principle, given stellar interior profiles obtained from stellar
simulations, our calculation presented in this work can be readily
applied to other stars. }, aiming at providing an update of the previous result~\cite{CAST:2007jps}:

\begin{equation}
\frac{d\Phi_{a}}{dE_{a}}=6.02\times10^{10}{\rm cm}^{-2}\text{s}^{-1}\text{keV}^{-1}g_{{\rm 10}}^{2}\overline{E}_{a}^{2.481}e^{-\overline{E}_{a}/1.205}\thinspace,\label{eq:-74}
\end{equation}
where $g_{{\rm 10}}\equiv g_{a\gamma}\times10^{10}\ \text{GeV}$ and
$\overline{E}_{a}\equiv E_{a}/\text{keV}$. Eq.~\eqref{eq:-74} was
first used by the CAST experiment~\cite{CAST:2007jps} and also
subsequently adopted by many solar axion searches at CDMS~\cite{CDMS:2009fba},
EDELWEISS~\cite{Armengaud:2013rta},  MAJORANA~\cite{Majorana:2022bse}.
 It is obtained by integrating the Primakoff production rate for
a solar model by Bahcall {\it et al} published in 2004~\cite{Bahcall:2004fg},
assuming massless axions. The solar model dependence is recently
examined by Ref.~\cite{Hoof:2021mld} and known to cause percent-level
variations. We note here that the corrections to the cross section
for massless axions can also be as large as $1\%$. Therefore, if
percent-level accuracy of the solar axion flux is required, Eq.~\eqref{eq:-74}
needs to be revised by taking both the solar model dependence and
the corrections to the cross section into account. For massive axions
with $m_{a}/E_{\gamma}\gtrsim0.1$, according to Fig.~\ref{fig:correction},
the finite-$m_{a}$ correction becomes more significant. Hence for
keV axions, it is necessary to include the finite-$m_{a}$ correction
in the calculation of the solar axion flux. 

In this work, we take an updated solar model  and apply our more
elaborated Primakoff cross section to recalculate the flux. In the
massless limit, our flux is very close to Eq.~\eqref{eq:-74} but
for keV axions it deviates significantly from the vacuum limit. Therefore,
our result would be important to solar axion search experiments using
crystal~\cite{SOLAX:1997lpz,CDMS:2009fba,Armengaud:2013rta,Majorana:2022bse}
or liquid xenon~\cite{Abe:2012ut,XENON100:2014csq,Gao:2020wer,Dent:2020jhf}
detectors, which are sensitive to deposit energies above a few keV.
 Note that in this work we assume that the axion-electron coupling
is absent, otherwise there is an additional contribution to the flux,
the so-called ABC flux in the literature~\cite{Redondo:2013wwa}.

\begin{figure}
\centering

\includegraphics[width=0.49\textwidth]{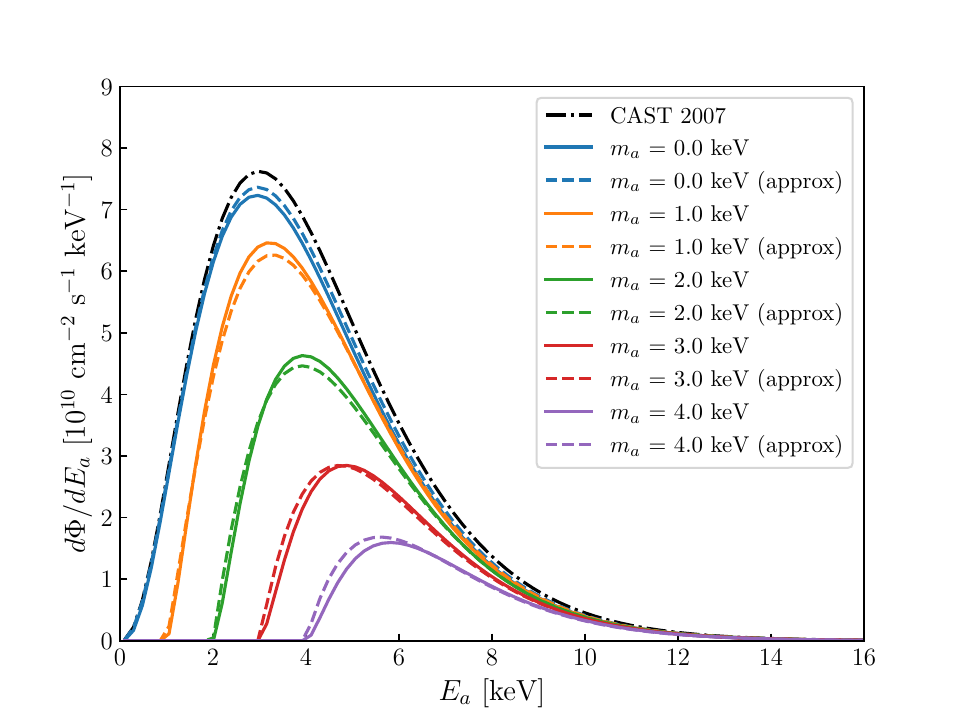}\raisebox{-0.075cm}{\includegraphics[width=0.51\textwidth]{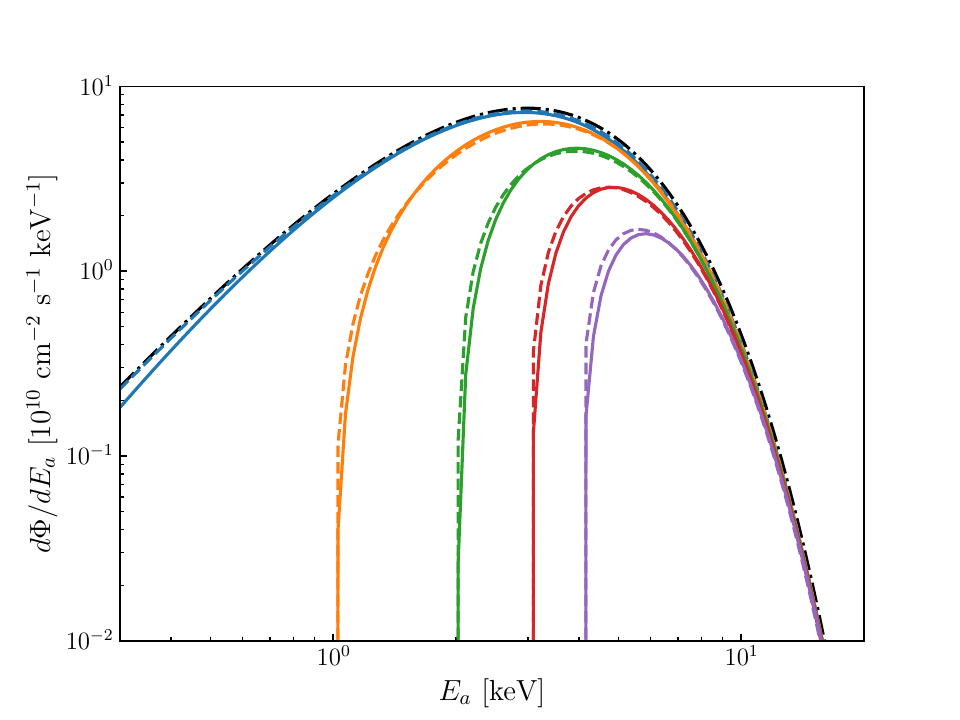}}\caption{Solar axion flux computed in this work. The solid lines are obtained
from our numerical calculation while the dashed lines are produced
using the approximate expression in Eq.~\eqref{eq:-84}. The black
dot-dashed line represents the CAST 2007 expression in Eq.~\eqref{eq:-74}.
The right panel represents the left in log scales, with the same color
code. \protect\label{fig:flux}}

\end{figure}

Let us first briefly review some basic formulae for the calculation
of solar axion flux. The probability of a photon scattering off a
medium particle ($X$) per unit time is $\sigma_{X}n_{X}$ where
$\sigma_{X}$ is the cross section of $\gamma$ with $X$ and $n_{X}$
is number density of $X$. We sum over all possible medium particles
so the total conversion rate of a photon into an axion reads:
\begin{equation}
\Gamma_{\gamma\to a}=\sum_{X}\sigma_{X}n_{X}\thinspace.\label{eq:-75}
\end{equation}
If we take  the simplified cross section in Eq.~\eqref{eq:-78},
$\Gamma_{\gamma\to a}$ can be written as
\begin{equation}
\Gamma_{\gamma\to a}=\frac{g_{a\gamma}^{2}\kappa_{s}^{2}T}{32\pi}\left[\left(1+\frac{\kappa_{s}^{2}}{4E_{\gamma}^{2}}\right)\log\left(1+\frac{4E_{\gamma}^{2}}{\kappa_{s}^{2}}\right)-1\right],\ \ m_{a}\to0\ \&\ m_{X}\to\infty\thinspace,\label{eq:-79}
\end{equation}
which agrees with Eq.~(5) in Ref.~\cite{CAST:2007jps}. For general
values of $m_{a}$ and $m_{X}$, we refer to Appendix~\ref{sec:Full-expression}
for the full expression of the cross section and use it in Eq.~\eqref{eq:-75}. 

With $\Gamma_{\gamma\to a}$ obtained, the solar axion flux can be
computed by
\begin{equation}
\frac{d\Phi_{a}}{dE_{a}}=\int_{0}^{R_{\astrosun}}dr\frac{r^{2}}{D_{\astrosun}^{2}}\cdot\frac{4\pi E_{a}^{2}}{(2\pi)^{3}}\cdot\frac{2}{e^{E_{a}/T}-1}\Gamma_{\gamma\to a}\thinspace,\label{eq:-83}
\end{equation}
where we have neglected the small difference between $E_{a}$ and
$E_{\gamma}$. The difference could be included by using the differential
cross section multiplied by the photon distribution function, and
integrating it with respect to $E_{\gamma}$. 

Using Eqs.~\eqref{eq:-75} and \eqref{eq:-83} with the cross section
in Appendix~\ref{sec:Full-expression}, we compute the solar axion
flux for $m_{a}\in\{0,\ 1,\ 2,\ 3,\ 4\}\ \text{keV}$ and present
the results in Fig.~\ref{fig:flux}. For the solar model, we take
B16-AGSS09met from Ref.~\cite{Vinyoles:2016djt}. 

As is shown in Fig.~\ref{fig:flux}, our massless limit (blue curve)
approximately agrees with the previous CAST 2007 result (black curve)
given by Eq.~\eqref{eq:-74}. Our result is slightly lower than the
previous, owing to solar model variations. The difference, however,
becomes rather significant for the axion mass at the keV scale, 
implying that Eq.~\eqref{eq:-74} should be modified when used for
keV axion searches. In practice, we find that the following expression
(shown as dashed lines in Fig.~\ref{fig:flux}) can approximately
fit our numerical result:

\begin{equation}
\frac{d\Phi_{a}}{dE_{a}}\approx5.94\times10^{10}{\rm cm}^{-2}\text{s}^{-1}\text{keV}^{-1}g_{{\rm 10}}^{2}\overline{E}_{a}^{2.49}e^{-\overline{E}_{a}/1.19}S\thinspace,\label{eq:-84}
\end{equation}
where $S$ is an suppression factor to account for the mass correction:
\begin{equation}
S=\begin{cases}
1-\left(m_{a}/E_{a}\right)^{1.67} & E_{a}>m_{a}\\
0 & E_{a}\leq m_{a}
\end{cases}\thinspace.\label{eq:-85}
\end{equation}
Note that Eq.~\eqref{eq:-84} should not be used in the strong Boltzmann
suppression regime since the parametric fit is performed within an
energy range up to $\sim10$ keV. For higher axion masses or energies,
the full numerical approach should be adopted to compute the Boltzmann
suppressed tail more accurately and, according to Ref.~\cite{Lucente:2022wai},
the on-shell photon coalescence ($\gamma\gamma\to a$) should be included.
This can be implemented by integrating $|{\cal M}_{\gamma\gamma\to a}|^{2}=g_{a\gamma}^{2}m_{a}^{4}/2$
with two photon distribution functions.

\section{Conclusions \protect\label{sec:Conclusions}}

The Primakoff process has a pivotal role in the production and detection
of axions in astrophysical environments and laboratories. Given the
on-going activities in axion searches, we revisit this process and
present a comprehensive calculation including various previously neglected
factors. In this work, we adopt a consistent QFT derivation with rigorous
treatment of the kinematics, and obtain Primakoff cross sections that
are valid in the full mass range allowed by the kinematics. Regarding
the photon-axion conversion in a magnetic background, we also obtain
a new result that remains valid for non-relativistic axions. Further,
we update the calculation of the solar axion flux, providing both
a simple and practically useful expression {[}see Eq.~\eqref{eq:-84}{]}
and full numerical access with the code publicly available at GitHub~\href{https://github.com/Fenyutanchan/Solar-Axion-Primakoff-Flux.git}{\faGithub}.
Compared to the previous widely-used one {[}see Eq.~\eqref{eq:-74}{]},
the most prominent difference in our new result is caused by the mass
correction, which becomes significant for keV or heavier axions. Therefore,
our work would be of particular importance to axion search experiments
utilizing crystal and liquid xenon detectors.

\begin{acknowledgments}
We acknowledge the use of \textsc{FeynCalc}~\cite{Mertig:1990an,Shtabovenko:2016sxi,Shtabovenko:2020gxv}
and \textsc{Package-X}\footnote{This package is available at \url{https://gitlab.com/mule-tools/package-x.git}.}~\cite{Patel:2015tea,Patel:2016fam}
in this work.  X.-J.~Xu is supported in part by the National Natural
Science Foundation of China under grant No.~12141501 and also by
the CAS Project for Young Scientists in Basic Research (YSBR-099).
Q.-f.~Wu is supported by the National Natural Science Foundation
of China under grant No.~12075251.
\end{acknowledgments}

\appendix

\section{The QFT approach to coherent scattering with background fields\protect\label{sec:QFT-coh}}

In this appendix, we show that the conversion between two particles
such as the photon and the axion in a background field can be consistently
derived in the QFT approach. Unlike the classical field approach which
treats the incoming photon as an electromagnetic field and the outgoing
axion as another oscillating field, the QFT approach treats incoming
and outgoing states as single-particle states. The key concept here
is that the incoming particle undergoes coherent scattering with the
background field and is thereby converted to another particle, as
we shall elucidate below.

We start by considering a toy model which contains two massive scalar
field $\phi_{1}$ and $\phi_{2}$, with different masses $m_{1}$
and $m_{2}$, coupled to a background scalar field $\Phi$ as follows:
\begin{equation}
{\cal L}=\sum_{i=1,2}\left(\frac{1}{2}\partial^{\mu}\phi_{i}\partial_{\mu}\phi_{i}-\frac{1}{2}m_{i}^{2}\phi_{i}^{2}\right)+\phi_{1}\phi_{2}\Phi^{2}.\label{eq:-44}
\end{equation}
The background field $\Phi$ has a nonzero expectation value only
within a slice of space that extends infinitely along the $y$ and
$z$ axis, but is bounded along the $x$ axis:
\begin{equation}
\langle\Phi^{2}\rangle=\begin{cases}
v^{2} & (a\leq x\leq b)\\
0 & (\text{otherwise})
\end{cases}\thinspace.\label{eq:-45}
\end{equation}

\begin{figure}
\centering

\includegraphics[width=0.5\textwidth]{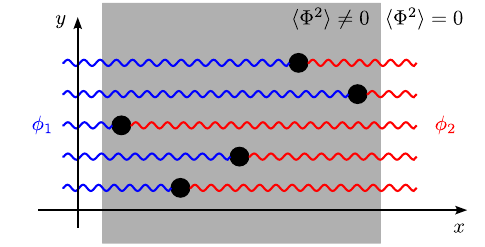}\caption{Conversion of a $\phi_{1}$ particle (blue) to a $\phi_{2}$ particle
(red) in the background field $\Phi$ (gray region) via the interaction
$\phi_{1}\phi_{2}\Phi^{2}$.  The conversion may  occur at any point
in the gray region and the overall effect can be viewed as coherent
scattering of $\phi_{1}$ with the background field, with all possible
scattering amplitudes added coherently.  \protect\label{fig:coh}}
\end{figure}

Now consider a $\phi_{1}$ particle moving along the $x$ axis and
going  through this slice of space. Due to the interaction $\phi_{1}\phi_{2}\Phi^{2}$,
there is a certain probability that the $\phi_{1}$ particle may be
converted to a $\phi_{2}$ particle, as illustrated by Fig.~\ref{fig:coh}. 

Denote the momenta of the $\phi_{1}$ and $\phi_{2}$ particles by
$p_{1}^{\mu}=(E_{1},\ \mathbf{p}_{1})$ and $p_{2}^{\mu}=(E_{2},\ \mathbf{p}_{2})$.
Since the background field is static and $m_{1}\neq m_{2}$,  we expect
that
\begin{align}
E_{1} & =E_{2}\thinspace,\label{eq:-46}\\
\mathbf{p}_{1} & \neq\mathbf{p}_{2}\thinspace.\label{eq:-47}
\end{align}
Eqs.~\eqref{eq:-46} and \eqref{eq:-47} will be derived later. Here
let us first discuss the physical implication of Eqs.~\eqref{eq:-46}
and \eqref{eq:-47}. The momentum difference $\mathbf{q}=\mathbf{p}_{1}-\mathbf{p}_{2}$
can be interpreted as the momentum transferred from $\phi_{1}$ to
the background. For coherent scattering, the inverse of the momentum
transfer needs to be much greater than the geometric dimension of
the target. As we are only concerned with the $x$ direction, this
implies 
\begin{equation}
|\mathbf{q}|\ll L^{-1}\thinspace,\label{eq:-48}
\end{equation}
where $|\mathbf{q}|=p_{1x}-\sqrt{p_{1x}^{2}-\delta m^{2}}$ with $\delta m^{2}\equiv m_{2}^{2}-m_{1}^{2}$,
and $L=b-a$ is the width of the gray region in Fig.~\ref{fig:coh}.
When Eq.~\eqref{eq:-48} is satisfied, full coherency is achieved,
implying that the scattering amplitude is proportional to $L$,  hence
the conversion probability  proportional to $L^{2}$: 
\begin{equation}
P_{\phi_{1}\to\phi_{2}}\propto L^{2}\thinspace.\label{eq:-49}
\end{equation}
As we will see, indeed the result derived later exhibits this feature
when Eq.~\eqref{eq:-48} is satisfied. 

Below we present the calculation. 

The amplitude of the $\phi_{1}$ particle being converted to $\phi_{2}$
reads:
\begin{equation}
i{\cal A}\equiv\langle2|e^{i\int H_{I}dt}|1\rangle\thinspace,\label{eq:-50}
\end{equation}
where $H_{I}$ is the interaction Hamiltonian  and the initial and
final states are defined as 
\begin{equation}
|i\rangle\equiv\int\frac{d\mathbf{k}_{i}}{(2\pi)^{3}}\frac{1}{\sqrt{2E_{\mathbf{k}_{i}}}}w_{i}(\mathbf{k}_{i})|\mathbf{k}_{i}\rangle\thinspace,\ \ |\mathbf{k}_{i}\rangle=\sqrt{2E_{\mathbf{k}_{i}}}a_{\mathbf{k}_{i}}^{\dagger}|0\rangle\thinspace,\ \ i=1,2\thinspace.\label{eq:-51}
\end{equation}
Note that $|\mathbf{k}_{i}\rangle$ is a single-particle state normalized
in the QFT convention, $\langle\mathbf{k}'_{i}|\mathbf{k}_{i}\rangle=2E_{\mathbf{k}_{i}}(2\pi)^{3}\delta^{3}\left(\mathbf{k}_{i}-\mathbf{k}'_{i}\right)$
which is Lorentz invariant, while $|i\rangle$ wraps $|\mathbf{k}_{i}\rangle$
with a normalized wavefunction $w_{i}(\mathbf{k}_{i})$ such that
\begin{equation}
\langle i|i\rangle=1\ \ \Leftrightarrow\int\frac{d\mathbf{k}_{i}}{(2\pi)^{3}}\left|w_{i}(\mathbf{k}_{i})\right|^{2}=1\thinspace.\label{eq:-52}
\end{equation}
The momentum distribution of $w_{i}(\mathbf{k}_{i})$ is assumed to
be centered around $\mathbf{p}_{i}$ with a very small but nonvanishing
width such that at macroscopic scales the two particles can be localized
in both momentum space and coordinate space---for further explanation,
see Appendix A of Ref.~\cite{Smirnov:2022sfo}. The specific form
of $w_{i}(\mathbf{k}_{i})$ is unimportant as long as it satisfies
the aforementioned requirements. 

The interaction Hamiltonian $H_{I}=\int\phi_{1}\phi_{2}\Phi^{2}d^{3}\mathbf{x}$
is assumed to be perturbative, which implies $i{\cal A}\approx\langle2|i\int H_{I}dt|1\rangle\approx i\int\langle2|\phi_{1}\phi_{2}\Phi^{2}|1\rangle d^{4}x$.

To proceed, we need to calculate \begin{equation}
{\cal C} \equiv
\langle
 \wick{
  \c1{\mathbf{k}_{2}}| \c2 \phi_{1} (x) \c1 \phi_{2} (x)\Phi^{2} (x)|\c2{\mathbf{k}_{1}}
  }
\rangle\,.
\label{eq:wick}
\end{equation}The quantized fields $\phi_{1}$ and $\phi_{2}$ (for simplicity,
we omit the subscripts when they are unimportant) are given by
\begin{equation}
\phi(x)=\int\frac{d\mathbf{k}}{(2\pi)^{3}}\frac{1}{\sqrt{2E_{\mathbf{k}}}}\left(a_{\mathbf{k}}e^{-ik\cdot x}+a_{\mathbf{k}}^{\dagger}e^{ik\cdot x}\right).\label{eq:-53}
\end{equation}

From Eq.~\eqref{eq:-53} and the definition of $|\mathbf{k}\rangle$
in Eq.~\eqref{eq:-51}, we obtain\begin{equation}
\wick{
\c1\phi(x)|\c1{\mathbf{k}} \rangle=e^{-ik\cdot x}|0\rangle\,.
}
\label{eq:wick2}
\end{equation}Hence, Eq.~\eqref{eq:wick} becomes
\begin{equation}
{\cal C}(x)=e^{ik_{2}\cdot x}e^{-ik_{1}\cdot x}\Phi^{2}(x)\thinspace.\label{eq:-63}
\end{equation}
Thus, the scattering amplitude is 
\begin{equation}
i{\cal A}=i\int d^{4}x\Phi^{2}(x)\int\frac{d^{3}\mathbf{k}_{1}}{(2\pi)^{3}}\frac{d\mathbf{k}_{2}}{(2\pi)^{3}}\frac{e^{i\left(k_{2}-k_{1}\right)\cdot x}}{\sqrt{2E_{\mathbf{k}_{1}}2E_{\mathbf{k}_{2}}}}w_{1}(\mathbf{k}_{1})w_{2}^{*}(\mathbf{k}_{2}).\label{eq:-54}
\end{equation}
Since the background field $\Phi(x)$ is time independent, we can
integrate out $dt$ with the time-dependent part of $e^{i\left(k_{2}-k_{1}\right)\cdot x}$,
giving rise to a Dirac delta function. Similar integration for the
spatial coordinates $y$ and $z$ can also be performed. Overall,
we obtain three Dirac delta functions:
\begin{align}
\int d^{4}x\Phi^{2}(x)e^{i\left(k_{2}-k_{1}\right)\cdot x} & =(2\pi)^{3}\delta\left(E_{\mathbf{k}_{1}}-E_{\mathbf{k}_{2}}\right)\delta\left(k_{1y}-k_{2y}\right)\delta\left(k_{1z}-k_{2z}\right)\nonumber \\
 & \times\tilde{\Phi}^{2}(k_{2x}-k_{1x})\thinspace,\label{eq:-55}
\end{align}
where $\tilde{\Phi}^{2}$ denotes the Fourier transform of $\Phi^{2}$:
\begin{equation}
\tilde{\Phi}^{2}(k)\equiv\int dx\Phi^{2}(x)e^{-ikx}\thinspace.\label{eq:-56}
\end{equation}
Eq.~\eqref{eq:-55} implies that the incoming and outgoing particles
should have the same energy as well as the same momentum in the $y$-$z$
plane. Since they have difference masses, their momenta along the
$x$ axis have to be different. 

Substituting Eq.~\eqref{eq:-55} into Eq.~\eqref{eq:-54}, it is straightforward
to integrate out $k_{2y}$ and $k_{2z}$ with two of the delta functions,
while the remaining one is to be integrated out by $k_{2x}$:
\begin{equation}
\int dk_{2x}\delta\left(E_{\mathbf{k}_{1}}-E_{\mathbf{k}_{2}}\right)\to\frac{E_{\mathbf{k}_{1}}}{\sqrt{E_{\mathbf{k}_{1}}^{2}-\tilde{m}_{2}^{2}}}=\frac{E_{\mathbf{k}_{1}}}{k_{2x}}\thinspace,\label{eq:-57}
\end{equation}
where $\tilde{m}_{2}^{2}\equiv m_{2}^{2}+k_{2y}^{2}+k_{2z}^{2}$. 

Note that when integrating $k_{2x}$ from $-\infty$ to $\infty$
in Eq.~\eqref{eq:-57}, there are two poles in $\delta\left(E_{\mathbf{k}_{1}}-E_{\mathbf{k}_{2}}\right)$,
one at $k_{2x}=\sqrt{E_{\mathbf{k}_{1}}^{2}-\tilde{m}_{2}^{2}}$ and
the other at $k_{2x}=-\sqrt{E_{\mathbf{k}_{1}}^{2}-\tilde{m}_{2}^{2}}$.
Both are physical solutions and can be understood intuitively, one
corresponding to coherent forward scattering and the other to coherent
backward scattering. We drop the latter since our derivation is focused
on coherent forward scattering. 

After integrating out all the three delta functions, we obtain
\begin{align}
i{\cal A} & =i\left.\int\frac{d^{3}\mathbf{k}_{1}}{(2\pi)^{3}}\tilde{\Phi}^{2}(k_{2x}-k_{1x})\frac{1}{\sqrt{2E_{\mathbf{k}_{1}}2E_{\mathbf{k}_{2}}}}\frac{E_{\mathbf{k}_{1}}}{k_{2x}}w_{1}(\mathbf{k}_{1})w_{2}^{*}(\mathbf{k}_{2})\right|_{\mathbf{k}_{2}\to(k_{2x},\thinspace k_{1y},\thinspace k_{1z})}\nonumber \\
 & =i\left.\int\frac{d^{3}\mathbf{k}_{1}}{(2\pi)^{3}}\frac{\tilde{\Phi}^{2}(k_{2x}-k_{1x})}{2k_{2x}}w_{1}(\mathbf{k}_{1})w_{2}^{*}(\mathbf{k}_{2})\right|_{\mathbf{k}_{2}\to(k_{2x},\thinspace k_{1y},\thinspace k_{1z})}\nonumber \\
 & =i\frac{\tilde{\Phi}^{2}\left(-|\mathbf{q}|\right)}{2p_{2x}}\left(\frac{p_{2x}}{p_{1x}}\right)^{1/2}\nonumber \\
 & =i\frac{\tilde{\Phi}^{2}\left(-|\mathbf{q}|\right)}{2\sqrt{p_{2x}p_{1x}}}\thinspace,\label{eq:-58}
\end{align}
where in the third step we have squeezed $w_{1}$ and $w_{2}$ to
their delta function limits {[}see the comments below Eq.~\eqref{eq:-52}{]}:
\begin{equation}
w_{1}(\mathbf{k}_{1})w_{2}^{*}(\mathbf{k}_{2})\to(2\pi)^{3}\sqrt{\delta^{3}(\mathbf{k}_{1}-\mathbf{p}_{1})}\sqrt{\delta^{3}(\mathbf{k}_{2}-\mathbf{p}_{2})}\thinspace.\label{eq:-59}
\end{equation}
Since the $y$ and $z$ components of $\mathbf{p}_{1}$ ($\mathbf{k}_{1}$)
and $\mathbf{p}_{2}$ ($\mathbf{k}_{2}$) are identical, we have $\delta(k_{2y}-p_{2y})=\delta(k_{1y}-p_{1y})$
and $\delta(k_{2z}-p_{2z})=\delta(k_{1z}-p_{1z})$. The part along
the $x$ axis however requires an additional treatment:
\begin{equation}
\delta(k_{2x}-p_{2x})\to\delta(k_{1x}-p_{1x})\left|\frac{\partial k_{2x}}{\partial k_{1x}}\right|^{-1}=\delta(k_{1x}-p_{1x})\frac{p_{2x}}{p_{1x}}\thinspace.\label{eq:-60}
\end{equation}

So for the toy model, our final result of the coherent forward scattering
amplitude is 
\begin{align}
{\cal A} & =\frac{1}{2\sqrt{p_{2x}p_{1x}}}\int dx\Phi^{2}(x)e^{i|\mathbf{q}|x}\thinspace\label{eq:-61}\\
 & =\frac{v^{2}}{2\sqrt{p_{2x}p_{1x}}}\cdot\frac{e^{i|\mathbf{q}|b}-e^{i|\mathbf{q}|a}}{|\mathbf{q}|}\thinspace,\label{eq:-62}
\end{align}
where in the second step we have used the profile of $\Phi^{2}(x)$
in Eq.~\eqref{eq:-45}. Note that Eq.~\eqref{eq:-61} can be generally
applied to arbitrary profiles of $\Phi^{2}(x)$, provided that it
only varies along the $x$-axis direction. 

The conversion probability is thus given by 
\begin{equation}
P_{\phi_{1}\to\phi_{2}}=|{\cal A}|^{2}=\frac{v^{4}}{p_{2x}p_{1x}|\mathbf{q}|^{2}}\sin^{2}\left(\frac{|\mathbf{q}|L}{2}\right)\thinspace.\label{eq:-64}
\end{equation}
This verifies our previous argument about  coherency that the probability
is proportional to $L^{2}$ when $|\mathbf{q}|$ is small---see Eq.~\eqref{eq:-49}.
Eq.~\eqref{eq:-64} is symmetric under the exchange of $1\leftrightarrow2$,
which implies that the inverse process has the same conversion probability,
$P_{\phi_{2}\to\phi_{1}}=P_{\phi_{1}\to\phi_{2}}$. 

The above calculation can be straightforwardly adapted to other scenarios
involving particles with internal degrees of freedom such as spins
or polarizations. For the photon-axion conversion in a magnetic field,
after proper treatments of the polarization of the  photon and the
orientation of the magnetic field, as has been addressed in Sec.~\ref{subsec:coh-mag-fields},
 the derivation is very similar if we replace $v^{2}\to g_{a\gamma}\langle B\rangle E_{\gamma}$.

\section{Full expression of the total cross section including Debye-H\"{u}ckel
screening\protect\label{sec:Full-expression}}

The total cross section of $\gamma+X\to a+X$, after including Debye-H\"{u}ckel
screening into Eq.~\eqref{eq:-19-1} and integrating out $T_{X}$,
is given as follows:
\begin{align}
\sigma_{{\rm tot}} & =\frac{\alpha g_{a\gamma}^{2}Q_{X}^{2}}{128E_{\gamma}^{2}m_{X}^{2}}\left[\left(q_{+}^{2}-q_{-}^{2}\right)\left(4m_{a}^{2}-8E_{\gamma}m_{X}-4m_{X}^{2}+q_{-}^{2}+q_{+}^{2}-2\kappa_{s}^{2}\right)\right.\nonumber \\
 & +2L_{2}\left(8E_{\gamma}^{2}m_{X}^{2}+m_{a}^{4}-4E_{\gamma}m_{a}^{2}m_{X}-4m_{a}^{2}m_{X}^{2}\right)\nonumber \\
 & +2L_{2}\left(2\kappa_{s}^{2}\left(2E_{\gamma}m_{X}+m_{X}^{2}-m_{a}^{2}\right)+2m_{a}^{4}m_{X}^{2}/\kappa_{s}^{2}+\kappa_{s}^{4}\right)\nonumber \\
 & \left.-4L_{1}m_{a}^{4}m_{X}^{2}/\kappa_{s}^{2}\right],\label{eq:-80}
\end{align}
where
\begin{equation}
q_{\pm}^{2}\equiv\frac{2E_{\gamma}^{2}m_{X}-m_{a}^{2}\left(E_{\gamma}+m_{X}\right)\pm E_{\gamma}\sqrt{4E_{\gamma}^{2}m_{X}^{2}-4m_{a}^{2}m_{X}\left(E_{\gamma}+m_{X}\right)+m_{a}^{4}}}{2E_{\gamma}+m_{X}}\thinspace,\label{eq:-81}
\end{equation}
\begin{equation}
L_{1}\equiv\log\left(\frac{q_{+}^{2}}{q_{-}^{2}}\right),\ \ L_{2}\equiv\log\left(\frac{q_{+}^{2}+\kappa_{s}^{2}}{q_{-}^{2}+\kappa_{s}^{2}}\right).\label{eq:-82}
\end{equation}

\bibliographystyle{JHEP}
\bibliography{ref}

\end{document}